\RequirePackage[2020-02-02]{latexrelease} 
\documentclass[twocolumn,showpacs,showkeys,preprintnumbers]{revtex4}
\usepackage{amsmath}
\usepackage{epsfig}
\usepackage{graphicx}
\begin{document}
	\title{%Bose-Einstein condensation of BCS Cooper pairs\\
%	Temperature-dependent gap as a promoter of two-step Bose-Einstein condensation\\
     % Two-step BEC coming from a temperature dependent energy gap
    BEC with two ground-state filling rates for BCS Cooper-pairs}
	% transici\'on
	%a la fase condensada de BE}

\author{J. J. Valencia$^1$ and M.A. Sol\'is$^2$ \\
	$^1$Universidad Aut\'onoma de la Ciudad de M\'exico, Plantel San Lorenzo Tezonco, \\
	Apdo. postal 09790, 09780 Ciudad de México, Mexico\\
	$^2$Instituto de F\'isica, Universidad Nacional Aut\'onoma de M\'exico, \\
	Apdo. postal 20-364, 01000 Ciudad de México, Mexico\\
juan.valencia@uacm.edu.mx, \hspace{1.0cm}masolis@fisica.unam.mx}

\date{Modified: \today / Compiled: \today}

%	\date{\today}

%\author{ J.J. Valencia and M.A. Sol\'{\i}s}
%\affiliation{Instituto de F\'{\i}sica, Apartado Postal 20-364, Universidad Nacional Aut\'onoma de M\'exico, M\'exico City, MEXICO }
%\date{Modified: \today / Compiled: \today}

\begin{abstract}
We report the effects on the thermodynamic properties of a 3D Bose gas caused by a temperature dependent energy gap between the ground state and the first excited state of the energy spectrum of the particles constituting the Bose gas which behaves like an ideal Bose gas when the gap is absent but whose properties are very different when it is present. Explicit formulae are given for the critical temperature, the condensate fraction, the internal energy and the isochoric specific heat, which are calculated for three different gaps which abruptly go to zero at a temperature $T_B$, as well as for the damped counterparts with a smoothed drop to zero. In particular, for the undamped BCS (Bardeen, Cooper and Schrieffer) gap it is observed that the Bose-Einstein condensation (BEC) critical temperature $T_c$ is equal to that of the ideal Bose gas $T_0$, for every $T_B \leq T_0$; surprisingly, the condensate fraction presents two different filling rates of the ground state, one at $T_c = T_0$ and another with a higher rate at $T_B < T_0$, suggesting something like a {\it two-step} BEC; also, its specific heat shows a finite jump at $T_c$ and a divergence at $T_B$ with a critical exponent $\alpha=1/2$, which is inherited from the divergence of the temperature derivative of the BCS gap at $T_B$.

\end{abstract}

\pacs{03.75.Fi; 05.30.Jp; 67.40.Kh}
\keywords{Bose-Einstein condensation; phase-transition; thermodynamic properties}

\maketitle

\section{Introduction}

Bose-Einstein condensation (BEC) is a physical phenomenon which has permeated many areas of physics, from condensed matter to cosmology \cite{Das}.
In particular, some increasingly accepted theories of superconductivity
 \cite{Lee, Tolmachev} consider Cooper pairs as composite bosons, an assumption that becomes more precise the smaller the correlation length of the pairs is, like in cuprates \cite{Hwang}. It is well known from BCS theory that Cooper pairs have a temperature-dependent energy gap between their ground and their first excited state energy, which promotes superconductivity. This gap begins with a finite value at $T= 0$ which decreases as a inverted half bell shape as the temperature increases, becoming zero at the superconducting critical temperature. 
On the other hand, some properties of superfluid helium four, such as its temperature dependent specific heat exponential behavior near $T=0$ was explained by London \cite{London}, suggesting that helium superfluidity is motivated by a BEC of the helium atoms with 
%and by introducing 
an artificial constant gap in their energy dispersion relation \cite{London2,japoneses}. Although the exponential behavior of the isochoric specific heat is reproduced for low temperatures, the BE critical temperature ($T_c$) increases beyond the BE critical temperature of an ideal Bose gas of helium atoms as the energy gap magnitude increases. This behavior is contrary to what is observed in helium, where the interaction among atoms reduces the superfluidity critical temperature below the 3D BEC critical temperature of an ideal Bose gas (IBG) when its particle mass and density are that of liquid helium four. Although many theoretical efforts \cite{gapjustification} have been made to justify the gap as a result of interactions among bosons, the gap not only has not showed up but there are arguments against its existence \cite{HP}. It is worth mentioning that %even though 
the critical temperature for an IBG plus a constant gap, like that proposed by London, has a BEC critical temperature which increases proportionally to the gap magnitude. However, when the gap decreases to zero as the temperature increases, we observed two possibilities: a) if the temperature $T_B$ at which the gap becomes zero, is less than or equal to $T_0$ (IBG critical temperature), the BEC critical temperature $T_c$ is equal to $T_0$, and  otherwise, b) if $T_B > T_0$ the temperature $T_c$ is greater than $T_0$ and depends on $T_B$ as well as on the gap shape. 
%
%temperature at which the gap becomes zero.b) the temperature critical is greater than T0 if the gap becomes zero at a temperature greater than T0. 

%when the gap decreases as the temperature increases, the BEC critical temperature is determined by the value of the temperature for which the gap becomes zero, i.e., with a temperature dependent gap London could have reproduced both the exponential behavior of the specific heat as a function of temperature as well as the critical temperature of superfluidity of helium four.
%
In this work we discuss the effects on the thermodynamic properties of a 3D Bose gas caused by a temperature dependent gap $\Delta (T)$ in the energy spectrum of the particles constituting the gas which behaves like an IBG without the gap. Although this problem seems to be an academic problem, there are evidences of  Bose gases with this type of dispersion relation like those emanating from imperfections in the networks where bosons are immersed \cite{juan} . 
%MAS SPINORS.

In Sec. II after defining our system, we calculate: a) the critical temperature as a function of the gap; b) the condensate fraction
and its derivative with respect to temperature which multiplied by the temperature and divided by the same condensed fraction, gives a filling rate.

  In Sec. III we give the following thermodynamic properties: the chemical potential, the internal energy, the isochoric specific heat as functions of temperature and the  jump height of the specific heat at $T_c$. In Sec. IV we apply the results of Sections II and III to three types of undamped and damped gaps but emphasizing the properties of the Bose gas with a BCS gap.
%design various types of temperature dependent gaps and apply them to the thermodynamic properties of our boson system.
In Sec. V we give our conclusions.

 \section{3D Bose gas with quadratic dispersion relation plus a temperature dependent gap}
 
Our system is a 3D infinite Bose gas of $N$ particles of mass $m$ whose dispersion relation energy as a function of its momentum magnitude $\hbar k$ is given by    
\begin{equation}
\varepsilon _{k}=\left\{
\begin{array}{c}
\varepsilon _{0}{\ \ \ \ \ \ \ \ \quad \ \quad \ \qquad \ \ \ \ \ \ \mbox{if}\ \ \  }k=0 \\
\varepsilon _{0}+\Delta(T) +\hbar^{2}k^{2}/2m{\ \ \ \mbox{if} \ \ \ }k>0%
\end{array}%
\right. ,  
\label{rdg}
\end{equation}
where $\Delta (T)$ is the energy gap between the ground state energy $\varepsilon_0$ and the first excited state energy $\varepsilon_0+\Delta(T)$. From here on we writte $\Delta_0 \equiv \Delta (T=0)$ and  $\Delta \equiv \Delta (T)$.

\subsection{BEC critical temperature}
For a finite temperature the particle number $N$ is distributed between the energy ground state and the excited states, i.e., $N=N_0(T)+ N_e(T)$ with 
\begin{equation}
N_{0}(T)={\frac{1}{e^{\beta \lbrack \mathbf{\varepsilon }_{0}-\mu (T)]}-1}}
\label{N0}
\end{equation}
the particles in the energy ground state and
\begin{equation}
N_e = \sum_{\mathbf{k\neq 0}}n_{k}  = \sum_{\mathbf{k\neq 0}}{\frac{1}{e^{\beta \lbrack \mathbf{\varepsilon }%
_{k}-\mu (T)]}-1}}   
\label{Ne1}
\end{equation}
the particles in the excited states,    
with $\beta \equiv 1/k_{B}T$, and $\mu (T)\leq \varepsilon _{0}$ the chemical potential.

In order to write the expression of $N_e$ in the thermodynamic limit, we convert the sum (\ref{Ne1}) into the integral
\begin{equation}
 N_e = \frac{1}{(2 \pi/L)^3} \int_{\mathbf{k\neq 0}}^{\infty}  {\frac{4 \pi k^2 dk}{e^{\beta (\varepsilon_{k}-\mu (T))}-1}},
 \label{Neint}
 \end{equation}
where $\varepsilon_{k} = \varepsilon_0 + \Delta (T) + \hbar^2 k^2 /2m $.
Defining $\xi \equiv \beta \hbar ^{2}k^{2}/2m$, 
%last equation becomes 
%from where we obtain the density of state $a(\xi)=L^{3}2m\xi/2\pi^{2}\beta\hbar^{2}$,
%
%\begin{eqnarray}
%	a(k)dk =\frac{L^{3}(2m/\hbar ^{2})^{3/2}\xi ^{1/2}d\xi}{4\pi ^{2}\beta^{3/2}}
%	\label{adk}
%\end{eqnarray}
%
%last 
%in this way the 
Eq. (\ref{Neint}) becomes
\begin{eqnarray}
N_e &=&\frac{L^{3}(2m/\hbar ^{2})^{3/2}}{4\pi ^{2}\beta^{3/2}}%
\int_{0}^{\infty }{\frac{\xi ^{1/2}d\xi }{e^{-\beta (\mu-\varepsilon_0-\Delta)} \ e^{\xi }-1}}  \nonumber \\
&{=}&\frac{L^{3}(2m/\hbar ^{2})^{3/2}}{4\pi ^{2}\beta^{3/2}}%
g_{3/2}(z_{1})\Gamma (3/2)
\label{Ne2}
\end{eqnarray}
with $z_{1} \equiv \exp[\beta(\mu - \varepsilon_0 - \Delta)]\leq 1$ and  $g_{3/2}(z_1)$ is the Bose function of order 3/2 \cite{Pathria}. 

%\subsection{BEC critical temperature}

The Bose Einstein critical temperature $T_c$ is the smaller temperature for which  
 $N_{0}(T_{c})/N \simeq 0$ and $N_e/N\simeq 1$, so
as a consequence $\mu(T)=\varepsilon_{0}$ for $T \leq T_c$. Therefore at $T_c$
\begin{equation}
N = \frac{L^{3}(2m/\hbar^{2})^{3/2}}{4\pi ^{2}\beta _{c}^{3/2}}%
g_{3/2}(z_{1c})\Gamma (3/2)
\label{Tc}
\end{equation}
with $z_{1c} \equiv \exp[-\beta _{c}\Delta(T_c)]\leq 1.$ 
For $\Delta_0 = 0 $, $z_{1c} = 1$ and we recover the critical temperature $T_{0}$ of an IBG via the relation
\begin{equation}
N = \frac{L^{3}(2m/\hbar^{2})^{3/2}}{4\pi ^{2}\beta _{0}^{3/2}}%
\zeta (3/2) \Gamma (3/2),
\label{T0}
\end{equation} 
with $\beta _{0}=1/k_BT_0$ and $T_0$ the BEC critical temperature for an IBG with a number density equal to that of our system. Note that in this case the dispersion relation for the Bose gas starts at $\varepsilon_0$ but has no effect on the critical temperature as this is a translation of the starting point of the particle energies. 

For $\Delta_0 \neq 0$, from the ratio between (\ref{Tc}) and (\ref{T0}),  
\begin{eqnarray}
1 &=&\frac{\beta _{0}^{3/2}g_{3/2}(z_{1c})}{\beta _{c}^{3/2}\zeta (3/2)}%
\qquad \nonumber\\
{\mbox{and} \qquad }T_{c} &=&\left( \frac{\zeta (3/2)}{g_{3/2}(z_{1c})}\right)
^{2/3}T_{0}
\label{eq:Tc/T0}
\end{eqnarray}
Since $z_{1c} \leq 1$, $g_{3/2}(z_{1c}) \leq \zeta(3/2)$ therefore $T_c \geq T_0$.

\subsection{Condensate fraction}

For $T<T_c$ the number of particles in the ground state $N_0(T)$ is comparable with the total number of particles $N$ and the condensate fraction $N_0(T)/ N$ is
\begin{equation}
	\frac{N_0(T)}{N}=1-\frac{N_e}{N}.
	\label{cf1}
\end{equation}
Sustituting (\ref{Ne2}) and (\ref{Tc}) in the last equation we obtain
\begin{equation}
	\frac{N_0(T)}{N}=1-\frac{T^{3/2}g_{3/2}(z_{1<})}{T_c^{3/2}g_{3/2}(z_{1c})}\xrightarrow[\Delta \to 0]{}  1- (T/T_0)^{3/2},
	\label{cf2}
\end{equation}
with $z_{1<} \equiv Exp[-\beta \Delta]$. When $T\leq T_c $,  $z_1  =z_{1<}$ and $z_{1c}=z_{1}(T_c)$. For $\Delta (T)= 0 $ we recover the condensate fraction for an IBG. 

Now to analyze how quickly the ground state is populated, we calculate the filling rate $(T/N_0)dN_0(T)/dT$. 
For $ T\leq T_c$ we derive $N_0$ given in Eq. (\ref{cf2}) with respect to the temperature  
\begin{equation}
\begin{split}
  	\frac{dN_0(T)}{dT}&=-\frac{NT^{1/2}}{T_c^{3/2}g_{3/2}(z_{1c})} \times \notag \\  &\left[\frac{3}{2}g_{3/2}(z_{1<})+g_{1/2}(z_{1<})\left(\frac{\Delta}{k_BT}-\frac{1}{k_B}\frac{d\Delta}{dT}\right)\right]
\end{split}
\end{equation}
and multiplying it by $T/N_0$, we obtain
  \begin{equation}
  	\frac{T}{N_0}\frac{dN_0(T)}{dT}=\frac{   \frac{3}{2}g_{3/2}(z_{1<})+g_{1/2}(z_{1<})\left(\frac{\Delta}{k_BT}-\frac{1}{k_B}\frac{d\Delta}{dT}\right)}{g_{3/2}(z_{1<})-\left(\frac{T_c}{T}\right)^{3/2}g_{3/2}(z_{1c})}.
  \end{equation}
  For $T>T_c$ we derive Eq. (\ref{N0})
  %
%  \begin{equation}
%  	N_0 (T) = \frac{1}{e^{\beta[\varepsilon_0-\mu(T)]}-1}
%  \end{equation}
%  and deriving
  \begin{equation}
  	\frac{dN_0}{dT} = \frac{e^{\beta[\varepsilon_0-\mu(T)]}}{\left[e^{\beta[\varepsilon_0-\mu(T)]}-1\right]^{2}}\beta \left[\frac{\varepsilon_0-\mu(T)}{T^{2}}+\frac{d\mu(T)}{dT}\right]
  \end{equation}
 which is multiplied by the coefficient $T/N_0$ to obtain 
\begin{equation}
  	\frac{T}{N_0}\frac{dN_0(T)}{dT}=\frac{\frac{\Delta}{k_BT}+\frac{3}{2}\frac{g_{3/2}(\exp[\beta(\mu - \varepsilon_0 - \Delta)])}{g_{1/2}(\exp[\beta(\mu - \varepsilon_0 - \Delta)])}}{e^{\beta(\mu-\varepsilon_0)}-1}.
\end{equation}
  The expression for the derivative of the chemical potential ${d\mu}/{dT}$ is given in Eq. (\ref{dmu}) of the next section.
% We will use of the expressions  

\section{Thermodynamic properties}

\subsection{Chemical potential}

Every thermodynamic properties of our system are chemical potential dependent.

For $T> T_c$ all bosons are in the excited states, i.e., $N=N_e$. Equating equations (\ref{Ne2}) and (\ref{T0}), and after canceling some factors, we obtain
\begin{equation}
	\frac{g_{3/2}(z_{1})}{\beta^{3/2}} = \frac{\zeta (3/2)}{\beta _{0}^{3/2}}.
\end{equation}
\noindent 
 Remembering that  $z_{1} = \exp[\beta(\mu - \varepsilon_0 - \Delta)]$, the last equation can be rewritten as
\begin{equation}
	(T/T_0)^{3/2} g_{3/2}(\exp[\beta(\mu - \varepsilon_0 - \Delta)]) = \zeta (3/2).
	\label{mu}
\end{equation}
This is an implicit equation for the chemical potential $\mu$ as a function of temperature, valid only for $T \geq T_c$, since for $T \leq T_c$  $\mu=\varepsilon_0$.

To calculate the specific heat we need the temperature derivative of the chemical potential, which we obtain differentiating expression (\ref{mu})
\begin{equation}
	\begin{split}
	\frac{1}{k_B}\frac{d\mu}{dT}=&\frac{1}{k_B}\frac{d\Delta}{dT}+\frac{\mu-\mathbf{\varepsilon }_{0}-\Delta}{k_BT} \\
	&-\frac{3}{2}\frac{g_{3/2}(\exp[\beta(\mu - \varepsilon_0 - \Delta)])}{g_{1/2}(\exp[\beta(\mu - \varepsilon_0 - \Delta)])}.
\label{dmu}
\end{split}
\end{equation}

\subsection{Internal energy}
The internal energy $U(L^{d},T)$ is given by the sum of the energies of the particles both in the ground and in the excited states, such that using  Eqs. (\ref{N0}) and (\ref{Ne1}) we get
\begin{equation}
	U(L^{d},T)={\frac{\varepsilon_{0}}{e^{\beta \lbrack \mathbf{\varepsilon }_{0}-\mu (T)]}-1}}
	+
	\sum_{\mathbf{k\neq 0}}{\frac{\varepsilon_{k}}{e^{\beta \lbrack \mathbf{\varepsilon }%
				_{k}-\mu (T)]}-1}} \ .  
	\label{Ie1}
\end{equation}

 In the thermodynamic limit, and using the energy espectrum Eq. (\ref{rdg}), we obtain
\begin{equation}
	\begin{split}
&	U(L^{d},T)={\frac{\varepsilon_{0}}{e^{\beta [\varepsilon_{0}-\mu (T)]}-1}} \\
	&+\frac{L^{3}4\pi}{(2\pi )^{3}}
	\int_{0^{+}}^{\infty }\frac{(\varepsilon _{0}+\Delta +\hbar^{2}k^{2}/2m)k^{2}dk}{e^{\beta{[\varepsilon _{0}+\Delta +\hbar^{2}k^{2}/2m-\mu (T)]}}-1}.  
\end{split}
	\label{Ie2}		
\end{equation}

Defining $\xi \equiv \beta \lbrack \hbar ^{2}k^{2}/2m]$,  $d\xi=(\beta \hbar ^{2}/2m)2kdk$ and $\xi ^{3/2}=(\beta \hbar ^{2}/2m)^{3/2}k^{3}$, the last equation becomes
\begin{equation}
	\begin{split}
	U(L^{d},T)&  =	\mathbf{\varepsilon }_{0}N_0+   \\
	& (\mathbf{\varepsilon }_{0}+\Delta)\frac{L^{3}(2m/\hbar ^{2})^{3/2}}{4\pi ^{2}\beta^{3/2}} 
	\int_{0}^{\infty }{\frac{\xi ^{1/2}d\xi }{z^{-1} \ e^{\xi }-1}}  \\
&	+\frac{L^{3}}{4\pi ^{2}}\frac{(2m/\hbar ^{2})^{3/2}}{\beta ^{5/2}}%
	\int_{0^{+}}^{\infty }\frac{\xi ^{3/2}d\xi }{z_{1}^{-1}e^{\xi }-1}.
	\end{split}
\label{UIB}
\end{equation}
In the right hand side of the last equation, the second term is the number of excited particles $N_e$ times ($\mathbf{\varepsilon }_{0}+\Delta$) (see Eq. (\ref{Ne2})). So  %Then, remembering that $N= N_0+N_e$, 
%we rewrite the internal energy as
%
\begin{equation}
	\begin {split}
	U(L^{d},T)&=	\mathbf{\varepsilon }_{0}N+\Delta N_e \\
	&+\frac{L^{3}}{4\pi ^{2}}\frac{(2m/\hbar ^{2})^{3/2}}{\beta ^{5/2}}%
	\Gamma(5/2)g_{5/2}(z_1),
	\end{split}
\label{eq:U3}
\end{equation}
where we have replaced the last integral of (\ref{UIB}) by its value $\Gamma(5/2)g_{5/2}(z_1)$ (see Appendix D \cite{Pathria}). Dividing it by $Nk_B T$, and taking the values of $N_e$ and $N$ given in Eqs. (\ref{Ne2}) and (\ref{Tc}) respectively, equation (\ref{eq:U3}) becomes
\begin{equation}
	\begin{split}
&	\frac{U(L^{d},T)}{Nk_B T}=\frac{\mathbf{\varepsilon }_{0}}{k_B T} \\
	& + \left[\frac{\Delta (T)}{k_B T}\frac{g_{3/2}(z_1)}{g_{3/2}(z_{1c})}  
	+\frac{3}{2}\frac{g_{5/2}(z_1)}{g_{3/2}(z_{1c})}\right]\left(\frac{T}{T_c}\right)^{3/2}.
	\label{UkBT}
	\end{split}
\end{equation}
%
%The first term suggests that the energy of the bosonic system increases due to the reference energy  $\mathbf{\varepsilon }_{0}$, while the second term suggests that the bosons in the exited states have to cross an energy $\Delta (T)$. 

%REVISAR LA DEDUCCION PORQUE LA EC. (21) ES U(T)/nk_B T 
\subsection{Isochoric specific heat}
The isochoric specific heat comes from deriving the internal energy (\ref{eq:U3}) with respect to the temperature, where we have used $N_e$ given in equation (\ref{Ne2}).  For $T \leq T_c $ we obtain 
\begin{eqnarray}
%	\hspace{-0.5cm} \begin{split}
 	&& \hspace{-0.5cm} \frac{C_V}{Nk_B} =
	\left[\left( \frac{\Delta^{2}}{(k_BT)^{2}}-\frac{\Delta}{k_B^{2}T}\frac{d\Delta}{dT}\right)g_{1/2}(z_{1<})+\frac{15}{4}g_{5/2}(z_{1<}) \right.  \nonumber \\
%	]\frac{\left(T/T_c\right)^{3/2}}{g_{3/2}(z_{1c})} \\
  &&+\left.\left(3 \frac{\Delta}{k_BT}-\frac{1}{2k_B}\frac{d\Delta}{dT}\right)g_{3/2}(z_{1<})\right]\frac{\left(T/T_c\right)^{3/2}}{g_{3/2}(z_{1c})}.
	\label{cvbajotc}
%	\end{split}
\end{eqnarray}	

	For $T \rightarrow 0$, $z_{1<} \rightarrow 0$, and using the approximation $g_{\sigma}(z_{1<}) \approx z_{1<}$ valid for $\sigma = 5/2, 3/2$ and $1/2$, the last Ec. (\ref{cvbajotc}) reduces to       
	\begin{eqnarray}
%	 \begin{split}
	&& \hspace{-0.2cm}	\frac{C_V}{Nk_B} =\left[ \frac{\Delta^{2}}{(k_BT)^{2}}-\frac{\Delta}{k_B^{2}T}\frac{d\Delta}{dT}+ \frac{3\Delta}{k_BT}-\frac{1}{2k_B}\frac{d\Delta}{dT}
		+\frac{15}{4}\right]   \nonumber \\
	&&	\times \frac{\mbox{Exp}[-\Delta/k_BT]\left(T/T_c\right)^{3/2}}{g_{3/2}(z_{1c})},
		\label{cvbajotcTpeque}
%	\end{split}
\end{eqnarray}
where we notice that for small temperatures the specific heat has an exponential behavior when we introduce a gap in the dispersion relation. In fact, for a constant gap $\Delta_0$
\begin{equation}
\hspace{-0.5cm}	\begin{split}
		\frac{C_V}{Nk_B}=\left[ \frac{\Delta_0^{2}}{(k_BT)^{2}}+\frac{3\Delta_0}{k_BT}
		+\frac{15}{4}\right]\frac{\exp[-\Delta_0/k_BT]\left(T/T_c\right)^{3/2}}{g_{3/2}(z_{1c})}
	\end{split}
	\label{CvbajotcTpequeDeltaConstante}
\end{equation}
\noindent
which decays exponentially as given by expression (28) of Ref. \cite{Martinez-Herrera} with $s=2$ and $d=3$, and
 %since the exponential decays faster than the polynomial 
%That is what is 
experimentally observed in helium four near absolute zero temperature. 

When $\Delta (T) = 0 $, i.e. the gap is absent,  Eq. (\ref{cvbajotcTpeque}) reduces to
\begin{equation}
	\begin{split}
		\frac{C_V}{Nk_B}= 
		\frac{15}{4}\frac{\left(T/T_c\right)^{3/2}}{\zeta(3/2)}
	\end{split}
	\label{CvbajotcTpequeDeltaCero}
\end{equation}
that is, $C_V$ is proportional to $T^{3/2}$ as expected for a 3D IBG. 

To obtain the specific heat for $T> T_c$ we derive the internal energy Eq. (\ref{eq:U3}) from where we obtain
\begin{equation}
	\begin{split}
		\frac{C_V}{Nk_B}=\frac{d}{dT}\left(\frac{U}{Nk_B}\right)= \frac{15}{4}\frac{g_{5/2}(z_{1})}{g_{3/2}(z_{1c})}\left(\frac{T}{T_c}\right)^{3/2} \\
		+\left[\frac{d\Delta (T)}{k_BdT}-\frac{9}{4}\frac{g_{3/2}(z_{1})}{g_{1/2}(z_{1})}\right]\frac{g_{3/2}(z_1)}{g_{3/2}(z_{1c})}\left(\frac{T}{T_c}\right)^{3/2} 
		\label{cvarribatc}
	\end{split}
\end{equation}
Note that for all temperature regimes, the specific heat is $\mathbf{\varepsilon }_{0}$ independent.  

On the other hand, since we have different expressions for the specific heat below and above $T_c$, we expect that for certain cases we obtain two different values for $C_V$ at the transition temperature. Let $C_V(T^{-}_c)$ be the specific heat from (\ref{cvbajotc}) evaluated at $T_c$ and $C_V(T^{+}_c)$ the value obtained by evaluating (\ref{cvarribatc}) at $T_c$. Thus, the specific heat discontinuity at $T_c$ is given by 
\begin{equation}
	\begin{split}
	  &	\frac{C_V(T^{-}_c)-C_V(T^{+}_c)}{Nk_B}= 
		\left[ \frac{\Delta^{2}}{(k_BT)^{2}}-\frac{\Delta}{k_B^{2}T}\frac{d\Delta}{dT} \right]_{T_c}\frac{g_{1/2}(z_{1c})}{g_{3/2}(z_{1c})} \\
		& +\left[\frac{3\Delta}{k_BT_c}- 
		 \frac{3}{2k_B}\frac{d\Delta}{dT}\right]_{T_c} +\frac{9}{4}\frac{g_{3/2}(z_{1c})}{g_{1/2}(z_{1c})}
		\label{SaltoDeCv}
	\end{split}
\end{equation}	
\noindent
when $\Delta =0$ we recover Eq. (14) of \cite{Aguilera-Navarro} for the case $d=3$ and $s=2$.
%
%\section\section{Condensed fractions}
%
%\section\section{Chemical potentials}
%
%\section\section{Specific heats}
%
%\section\section{Entropies}

\begin{figure}[htb]
	\vspace{-1.5cm}
%	\centerline{\epsfig{file=TresGapsSinFermi.pdf,height=3.8in,width=3.3in}}
\centerline{\epsfig{file=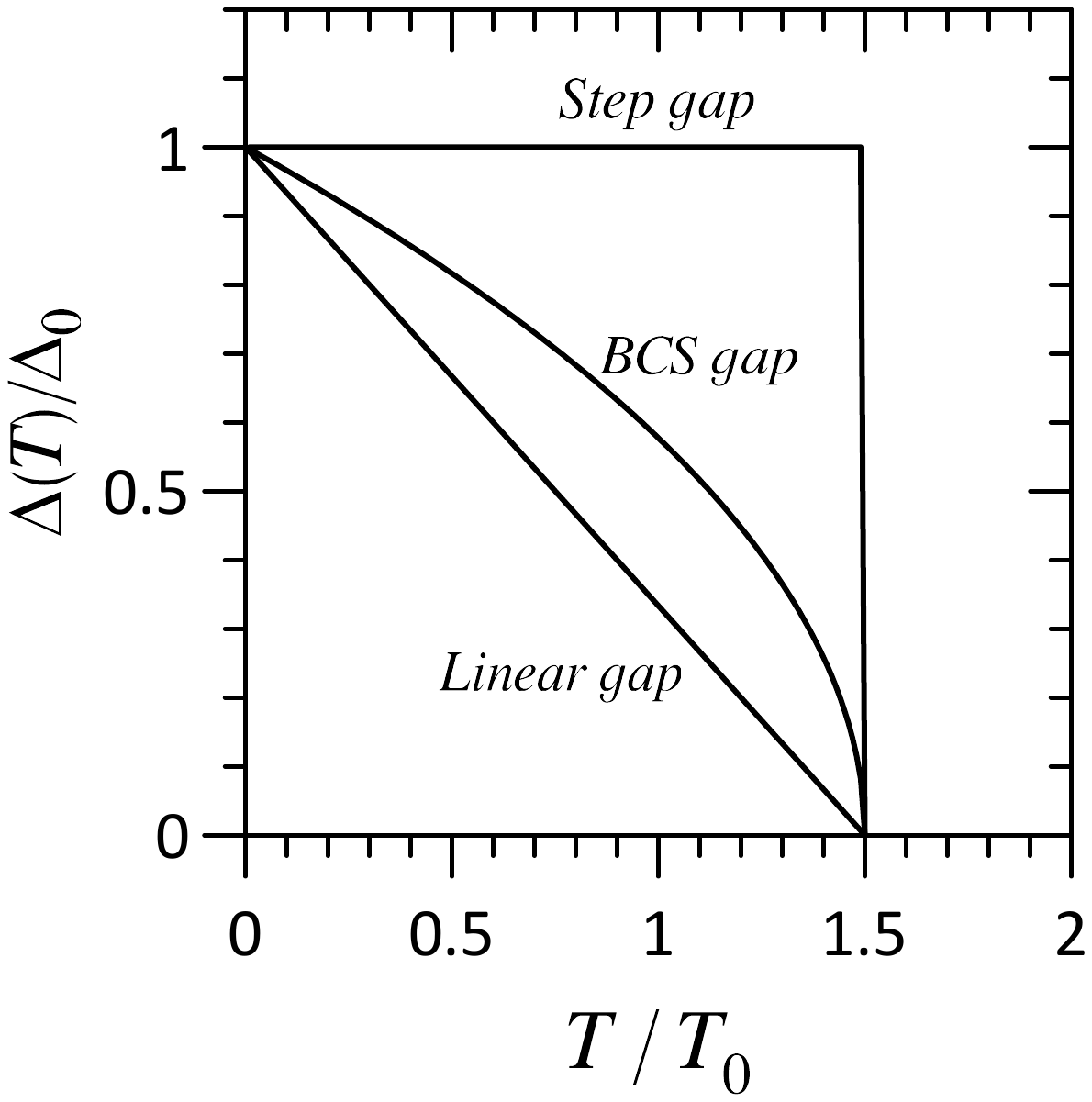,height=3.8in,width=3.3in}}
\vspace{-1.3cm}
	\caption{Examples of three temperature-dependent gaps studied here with  $\Delta_0=k_BT_0$ and $T_B=1.5T_0$. %Subscripts $L$, $S$ and $B$ mean Linear, Step and BCS, respectively.
	 }
	\label{fig:TresGapsSinFermi}
\end{figure}
\begin{figure}[htb]
	\vspace{-1.5cm}
	\centerline{\epsfig{file=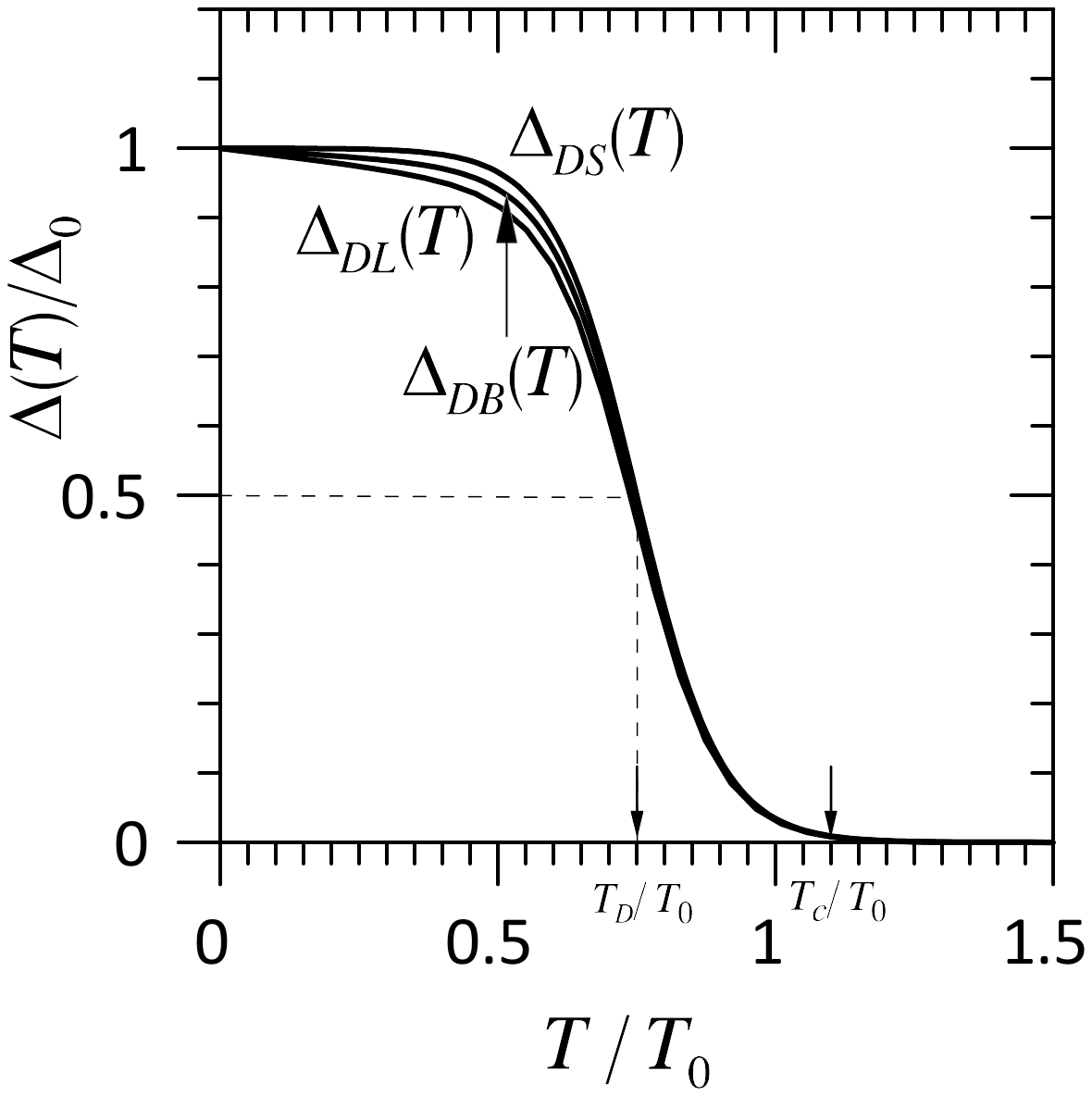,height=3.8in,width=3.3in}}
	\vspace{-1.3cm}
	\caption{The same gaps of Fig. \ref{fig:TresGapsSinFermi} but now multiplied by the damping function $f(T)=1/(\exp[b(T/T_D - 1)] + 1)$. For these  cases $\Delta_0 = k_BT_0$, $T_B = 10 \, T_0$, $b = 10$, $T_D = 0.75 \, T_0$. In particular for $\Delta_{DB}$ the resulting $T_c = 1.09 \, T_0$.  Subscripts $DL$, $DS$ and $DB$ mean Damped Linear, Damped Step and Damped BCS, respectively.}
	\label{fig:TresGaps}
\end{figure}
\section{Applications}
\subsection{Gap Types}
Here we calculate some thermodynamic properties with the expressions  given above for a Bose gas whose particles have an energy spectrum with one of six different temperature-dependent gaps. We begin by analyzing three  gaps which decrese with temperature, becoming zero above the temperature $T_B$, i.e., 
\begin{equation}
	\Delta (T) =\left\{
	\begin{array}{c}
		g(T)\ \ \ \ \ \ \ \ { \quad  \mbox{if}\ \ \  }T < T_{B} \\
		\quad 0 \quad \quad {\    \, \,\, \quad    \ \ \mbox{if} \ \ \ }T \geq T_{B}%
	\end{array}%
	\right.   
	\label{gapTB}
\end{equation}
where $g(T)=\Delta_0(1 - T/T_B)^{1/2}$ (BCS gap), $g(T)=\Delta_0(1-T/T_B)$ (Linear gap) or $g(T)=\Delta_0$ (Step gap).
%The para$T_B$ parameter is the temperature in which the gap becomes zero, this is not necessarily the critical temperature since there is a gap when there are pairs of cooper and above the critical temperature there may also be paired electrons forming which we are treating as bosons.
 In Fig. \ref{fig:TresGapsSinFermi} we show these three gaps for $T_B = 1.5 T_0$ and $\Delta_0 = k_B T_0$.
 % These are $ g(T)=\Delta_0(1 - T/T_B)^{1/2}$ (BCS gap), $g(T)=\Delta_0(1-T/T_B)$ (linear gap) and $g(T)=\Delta_0$ (step gap).

%
%\begin{equation}
%	\Delta (T) =\left\{
%	\begin{array}{c}
%		\Delta_0(1 - T/T_{B}){\ \ \ \ \ \ \ \ \quad  \mbox{if}\ \ \  }T\leq T_{B} \\
%		\quad 0{\  \qquad \qquad   \qquad \qquad \, \, \, \, \ \ \mbox{if} \ \ \ }T>T_{B}%
%	\end{array}%
%	\right. ,  
%	\label{bcsgap}
%\end{equation}
%
%\begin{equation}
%	\Delta (T) =\left\{
%	\begin{array}{c}
%		\Delta_0(1 - T/T_{B})^{1/2}{\ \ \ \ \ \ \ \ \quad  \mbox{if}\ \ \  }T\leq T_{B} \\
%		\quad 0{\  \qquad \qquad   \qquad \qquad \, \, \, \, \ \ \mbox{if} \ \ \ }T>T_{B}%
%	\end{array}%
%	\right. ,  
%	\label{bcsgap}
%\end{equation}

Although the BCS Theory suggests an abrupt creation of many Cooper pairs at and below the superconducting critical temperature, it is expected \cite{seudogap} that at a higher pseudo-gap  temperature pairs will begin to form, increasing in number as the system cools until it reaches a critical density to develop superconductivity.
 %which we assume is related to a BEC of Cooper pairs. 
 As an alternative to explore the effect of Cooper pairs creation above the superconducting critical temperature on the gap shape, we   
 %before to reach the critical density for a BEC 
 % with a exponential decreasing tail. %for temperatures larger than $T_d$. 
multiply the BCS gap, as well as the other two gap functions mentioned above, by the damping function $f(T) =1/(\exp[b(T/T_D - 1)] + 1)$, where $T_D$ is simply a temperature at which the damping function $f(T_D)=1/2$, and $b$ a large number to assure that when $T << T_D$ we recover the corresponding undamped gap function $g(T)$. Then, the expressions for the three damped gaps are
\begin{equation}
	\Delta_a (T) = \frac{g(T)}{\exp[b(T/T_D - 1)] + 1}
	\label{eq:Deltan}
\end{equation}
where
 %we call $T_D$ the damping temperature and 
  $a = DB, DL$ and $DS$, which mean damped BCS gap, damped Linear gap and damped Step gap, respectively. 
%
 %($DB\equiv $ damping BCS gap) when we do  $ g(T)=\Delta_0(1 - T/T_B)^{1/2}$, $a=DL$ ($DL\equiv $ damping linear gap) when  $ g(T)=\Delta_0(1 - T/T_B)$ and $a = DS$ ($DS\equiv $ damping step gap) when $g(T)=\Delta_0$. 
% 
 These damped gaps are plotted in Fig. \ref{fig:TresGaps} where, unlike Fig. \ref{fig:TresGapsSinFermi}, gaps have a smooth decay close to $T_0$ showing finite values of their derivatives, and going smoothly to zero at $T_B$. The parameters for the three damped gaps in Fig. \ref{fig:TresGaps} are $\Delta_0 = k_BT_0$, $T_B = 10 \, T_0$, $b = 10$, $T_D = 0.75 \, T_0$. For this $T_D$ the Bose gas with the gap  $\Delta_{DB}$  has a critical temperature $T_c = 1.09 \, T_0$ where the gap is different from zero but very small since its value is $0.0088 \ \Delta_0$. A similar smooth gap decay near $T_c$ is reported in Ref. \cite{Dougherty} for gaps of Nb, Ta, Pb, Hg, Sn, In, Al, Ga, Zn and Cd obtained from a thermodynamic analysis of experimental data.

\subsection{Gap effect on the BEC critical temperature}

\begin{figure}[htb]
	\vspace{-1.6cm}
	\centerline{\epsfig{file=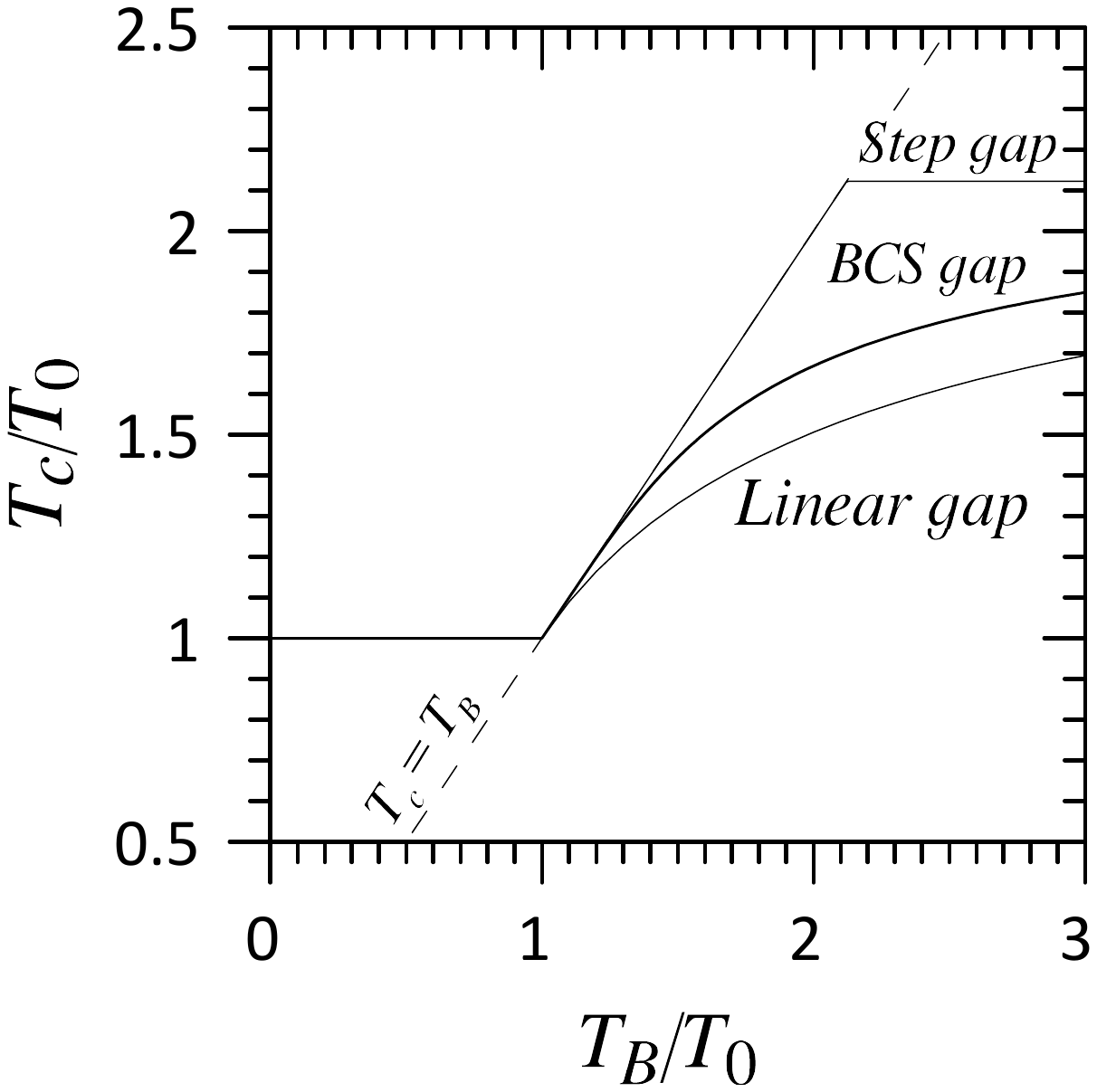,height=3.8in,width=3.3in}}
	\vspace{-0.8cm}
	\caption{Bose-Einstein critical temperature $T_c/T_0$ as function of $T_B/T_0$ for the three undamped gaps. Dashed line is $T_c=T_B$. }
	%	When our temperature dependent gaps are zero after $T_B$ (step gap, BCS gap and linear gap cases) we have $T_c$ tends to the constant gap case for large $T_B$. For small values of $T_B$ the temperature $T_c$ is $T_0$. When the gap is different from zero beyond $T_B$ (BCS-Constant gap case) we have that for small $T_B$ the critical temperature depends on the value of the gap after $T_B$. For large $T_B$, the critical temperature depends on the value that the gap takes before $T_B$.}
\label{fig:TcvsTBConstantBCSAndLinearGap}
\end{figure}
\begin{figure}[htb]
	\vspace{-2.1cm}
	\centerline{\epsfig{file=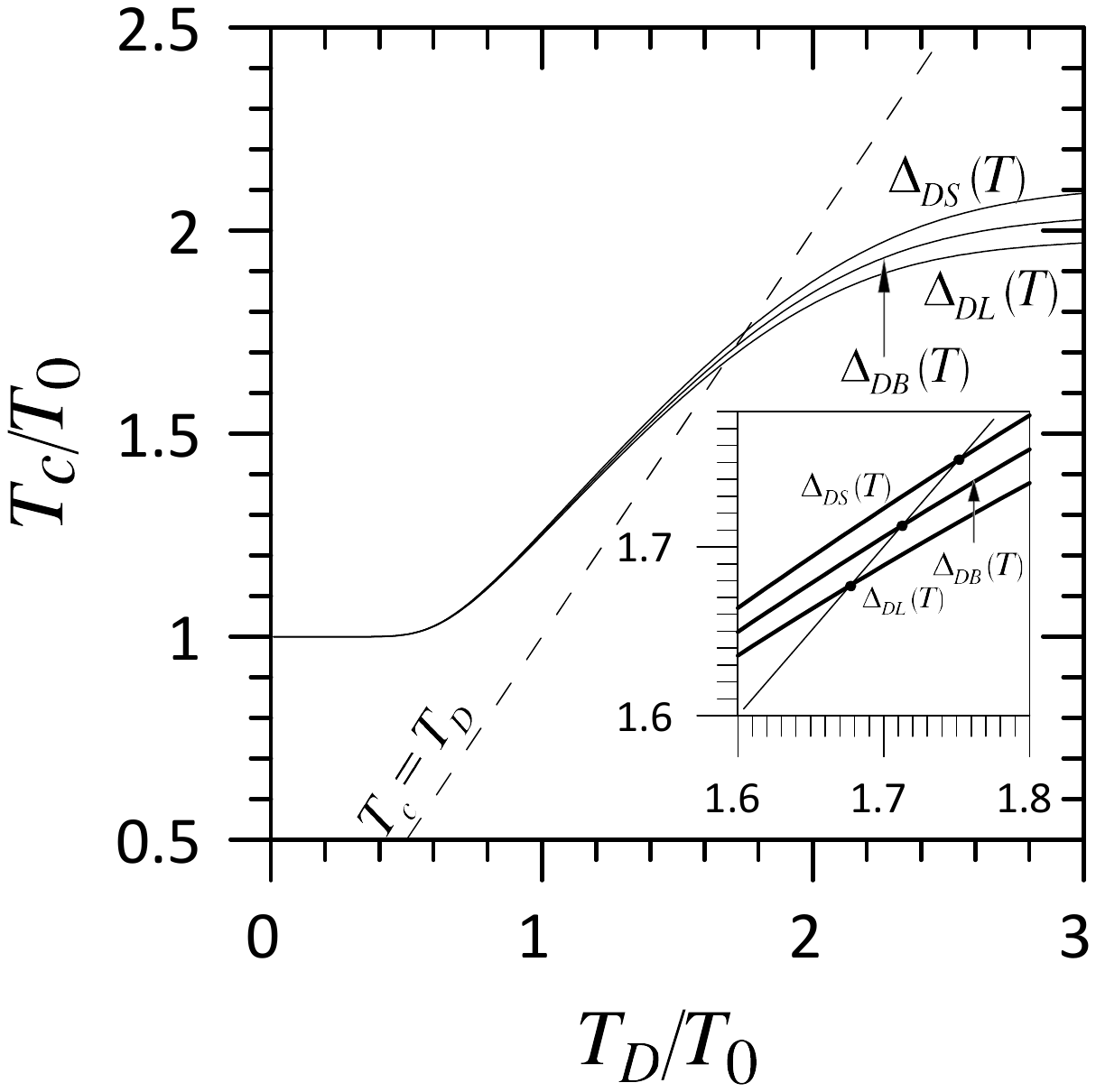,height=3.8in,width=3.3in}}
	\vspace{-0.7cm}
	\caption{Bose-Einstein critical temperature $T_c/T_0$ as a function of the damping temperature $T_D/T_0$ using the tree damped gaps of Eq. (\ref{eq:Deltan}), where $\Delta_0 = k_BT_0$, $T_B = 10 \, T_0$, and $b = 10$. Dashed line is $T_c=T_D$. In the inset we show the $T_D$ temperatures for which $T_D = T_C$.}
	\label{TcvsTdD1D2D3Gap}
\end{figure}
%London proposed that superfluid helium could be explained as a Bose-Einstein condensation. With this theory he calculated the critical temperature for superfluid helium, the result was 3.2 K. It is very close to the temperature at which helium four becomes superfluid, which is 2.17K. However, at temperatures close to absolute zero temperature, helium four grows exponentially and the growth of Bose gas is proportional to $T^{3/2}$. By introducing a constant gap in the dispersion relationship, he was able to reproduce the exponential behavior in Bose gas.  However, when the gap was introduced, the critical temperature grew with the size of the gap. If $\Delta=k_BT_0$ we obtain $T_c = 2.12 T_0$, the critical temperature is more than twice the calculated critical temperature without a gap, $T_0$. But if we introduce a temperature dependent gap, we can manipulate the critical temperature. 

In Fig. \ref{fig:TcvsTBConstantBCSAndLinearGap} we show the critical temperatures  as functions of $T_B/T_0$ using the undamped gaps described by Eq. (\ref{gapTB}) for the three forms of $g(T)$ (the BCS, the Linear and the Step gap). For all three gaps, regardless of their shapes,  we find that when  $T_B \leq T_0$ the BEC critical temperature $T_c = T_0$. When $T_B > T_0$ %and close to $T_0$,
the BEC critical temperature $T_c > T_0$ and gap shape dependent, except
%. This gap shape dependence of $T_c$ desappears 
when $T_B  >> T_0$, where the critical temperatures approach that for a constant gap $\Delta_0$ for every temperature.  
% ButThis is because for small $T_B$ the predominant behavior of the gap is zero, since its nonzero value predominates only in a small range of temperatures, from zero to $T_B$. Moreover, if we had a nonzero gap after $T_B$, this gap would determine the value of $T_c$ for small $T_B$ and could increase beyond $T_0$. This is demonstrated in the next subsection.

%In Fig. \ref{fig:TresGapsSinFermi} we show the BCS, Step and Linear gaps,
% proposed by the BCS theory, 
% which abruptly change from a non-zero value to zero when passing through the temperature $T_B$.
 We note that for the three gaps their temperature derivatives at $T_B$ have a jump,  which becomes infinite for the BCS gap. 
 %The same happens with the other two gaps shown in said figure, the step gap and the linear gap.
 % If we want to model a natural phenomenon without divergencies, we must eliminate infinities from our equations.
   In order to avoid a sudden drop to zero of the gap, which causes infinities in the thermodynamic properties,  
    %have a smooth passage from a non-zero gap to a gap equal to zero, 
    we multiply the undamped gaps by a damping function $f(T)$
   %
       %change the temperature $T_B$ by a damping temperature $T_D$ 
  as defined in Eq. (\ref{eq:Deltan}).
  %These damped gaps are shown in Fig. \ref{fig:TresGaps}.
  %\ref{TcvsTdD1D2D3Gap}.
   %which is similar to Fig. \ref{fig:TcvsTBConstantBCSAndLinearGap}. 
  
   In  Fig. \ref{TcvsTdD1D2D3Gap} we plot $T_c/T_0$, now as a function of the damping temperature $T_D/T_0$ 
   %instead of $T_B/T_0$ as was done in Fig. \ref{fig:TcvsTBConstantBCSAndLinearGap}. For this we 
   using the gaps $\Delta_{DB}(T)$, $\Delta_{DL}(T)$ and $\Delta_{DS}(T)$ shown in Fig. \ref{fig:TresGaps}. 
   Although Figs. \ref{fig:TcvsTBConstantBCSAndLinearGap} and \ref{TcvsTdD1D2D3Gap} have a similar behavior, their main differences are that using the undamped gaps we observe an abrupt change in the slopes of the critical temperature curves in $T_B/T_0 =1$, while when we use the damped gaps the critical temperature curves smooth for all temperatures $T_D/T_0$.
   
   %Another thing to mention is that 
  For each damped gap there is a temperature for which $T_D$ goes from being less than $T_c$ to being greater. In the inset of Fig. \ref{TcvsTdD1D2D3Gap} we show the mentioned temperatures: for the $\Delta_{DL}(T)$ gap, $T_D=1.68T_0$; for $\Delta_{DB}(T)$, $T_D=1.71T_0$; and in the case of $\Delta_{DS}(T)$ $T_D=1.75T_0$.
   
%   The figures \ref{fig:TcvsTBConstantBCSAndLinearGap} and \ref{TcvsTdD1D2D3Gap} CUALES? are very similar. The main difference is that in the first one we have an abrupt change from $Tc= T_0$ to $Tc> T_0$ in $T_B=T_0$, while in figure \ref{TcvsTdD1D2D3Gap} said change is smooth in $T_D=T_0$.  REVISAR REDACCION.

\bigskip
\subsection{Gap effect on the condensate fraction}

\begin{figure}[htb]
	\vspace{-1.4cm}
	\centerline{\epsfig{file=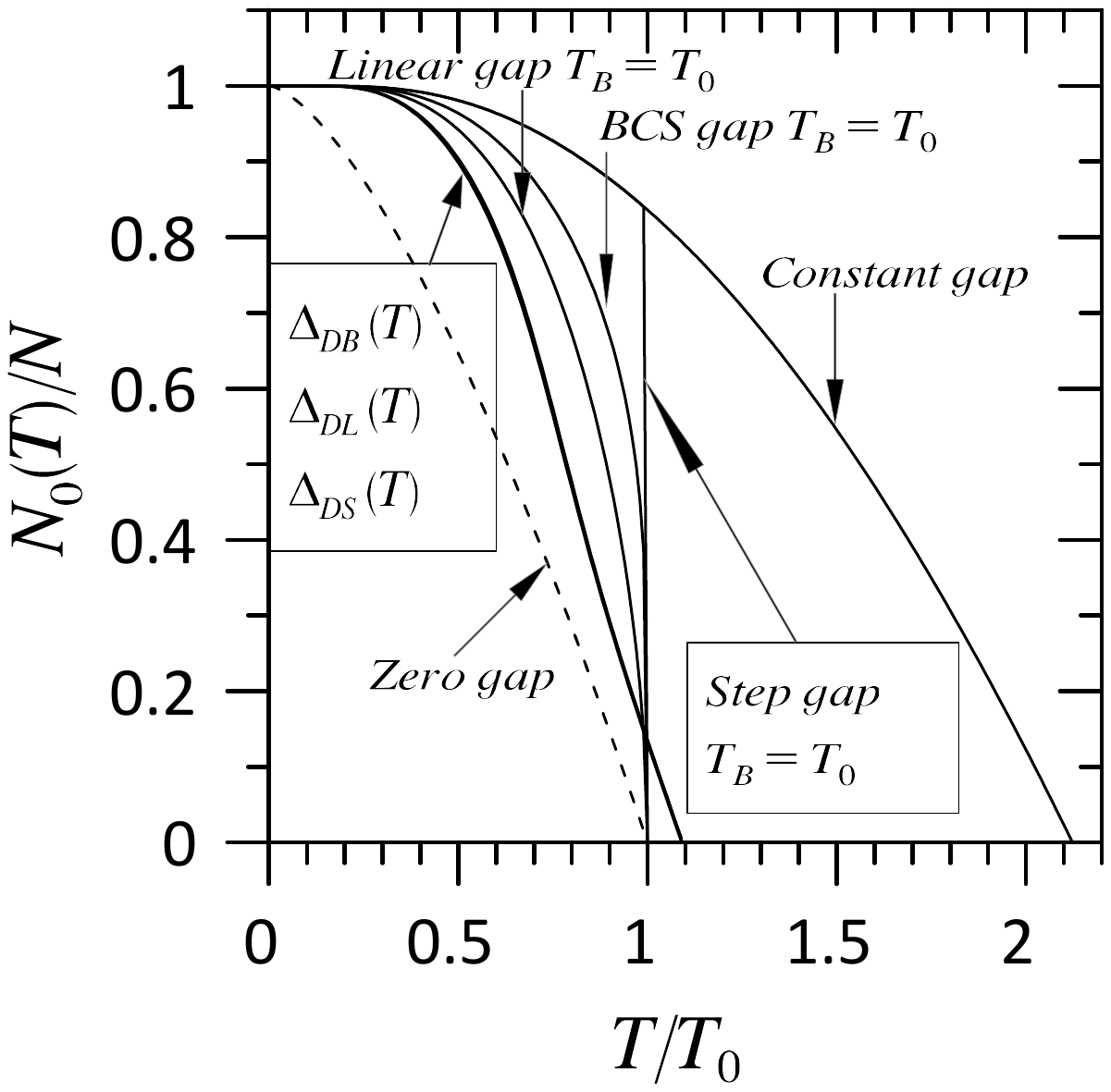,height=3.8in,width=3.3in}}
	\vspace{-1.3cm}
	\caption{Condensate fractions: the dashed lines correspond to the zero gap (IBG) case; for BCS, step and linear gaps $T_B =T_0$ was taken; for $\Delta_{DB}(T)$, $\Delta_{DL}(T)$, and $\Delta_{DS}(T)$ gaps the values $T_B=10\, T_0$, $b=10$, and $T_D=0.75\, T_0$ were used. For all cases $\Delta_0=k_BT_0$.}
	\label{fig:FcTresGaps}
\end{figure}
\begin{figure}[htb]
	\vspace{-1.2cm}
	\centerline{\epsfig{file=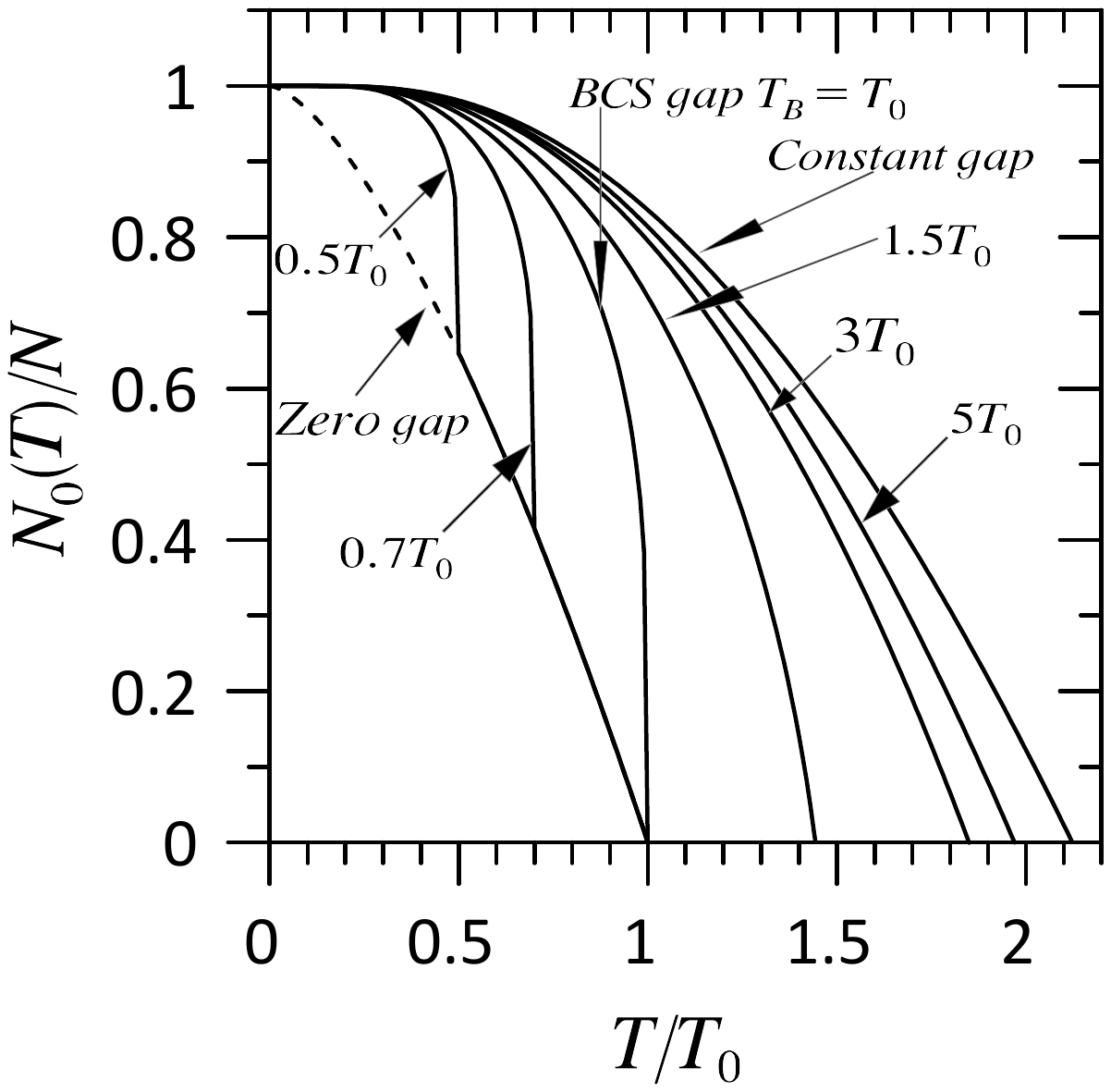,
			height=3.8in, width=3.3in}}
	\vspace{-1.3cm}
	\caption{Condensate fraction for BCS gap using $T_B =0.5T_0$, $T_B =0.7T_0$, $T_B =T_0$, $T_B =3T_0$, and $T_B =5T_0$ compared to the condensate fraction for zero and constant gap.}
	\label{fig:FcBCSGaps}
\end{figure}
 In Fig.  \ref{fig:FcTresGaps} we show the condensate fractions (CFs) as functions of temperature for the linear, BCS, and step undamped gaps; for all three gaps $\Delta_0 = k_BT_0$ and $T_B =T_0$. We note that from these three gaps, the linear one decreases faster with temperature (see Fig. \ref{fig:TresGapsSinFermi}) as well as its CF. These CFs are shown together with that of the zero gap (dashed line)(IBG) and that of the constant gap.   
 % In the figure we see that with this gap the fraction of the condensate decreases more rapidly with temperature. This is not surprising since the gap is proportional to the number of pairs; and therefore, the gap is proportional to the number of pairs in the condensate state. Note that by means of the constant gap the condensate fraction decreases more slowly compared to the other gaps.
%
%
Also,  in Fig. \ref{fig:FcTresGaps} are shown the CFs for the damped gaps $\Delta_{DB}(T)$, $\Delta_{DL}(T)$, and $\Delta_{DS}(T)$, which merge into one curve from zero to the critical temperature $T_c/T_0 = 1.09$, since the three gaps are very similar (see Fig. \ref{fig:TresGaps}). We note that the CFs do not reach zero with an  infinite slope as in the undamped BCS case. 
%; the $\Delta_{DB}(T)$, $\Delta_{DL}(T)$, and $\Delta_{DS}(T)$ gaps smooth the decay of the condensate fraction.  
\begin{figure}[htb]
	\vspace{-1.35cm}
	\centerline{\epsfig{file=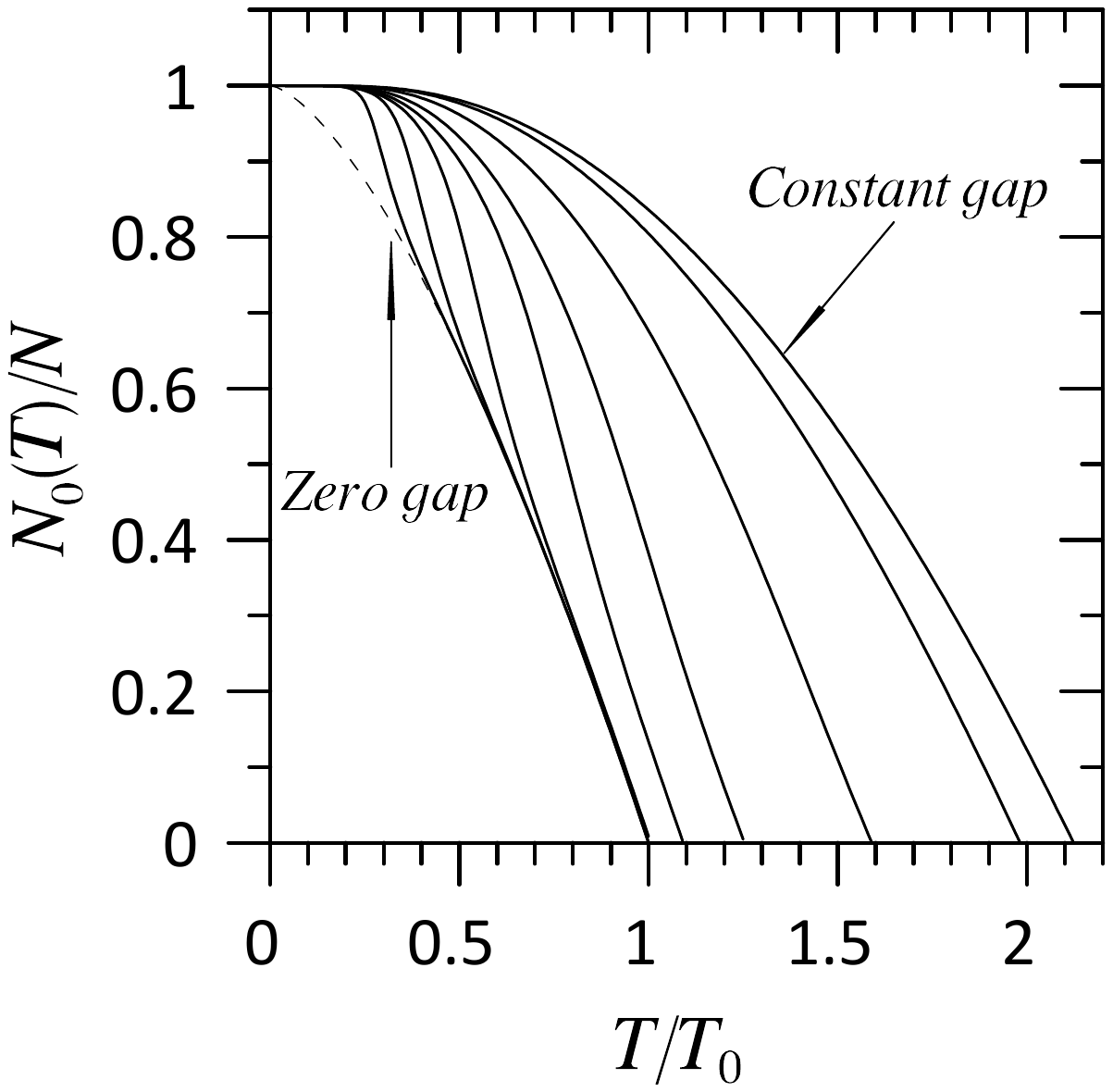,height=3.8in,width=3.3in}}
	\vspace{-1.4cm}
	\caption{Condensate fraction calculated: with zero gap (dashed line), using $\Delta_{DB}(T)$ gap with $b=10$, $T_B=10T_0$ and several values of $T_D$, from left to right: $T_D=0.25 T_0$, $T_D=0.35 T_0$, $T_D=0.50T_0$, $T_D=0.75 T_0$, $T_D= T_0$, $T_D=1.50 T_0$, $T_D=2.50 T_0$. And finally, the condensate fraction calculated with the constant gap $\Delta_0=k_BT_0$.}
	\label{fig:FcGapDB}
\end{figure}

In Fig. \ref{fig:FcBCSGaps} we show the CFs using the BCS undamped gap for several values of $T_B$ which we compare with the CFs without gap (IBG) and with a constant gap. For every $T_B < T_0$, the BEC critical temperature is $T_0$, while in the interval of temperatures $T_B < T  < T_0$, the CF curves merge with that of the IBG 
%precisely at the temperature 
but for $T \leq  T_B$ the CF magnitude increases rapidly until it reaches its maximum value $N_0/N = 1$ at $T=0$, showing a divergent temperature derivative at $T_B$ and two different filling rates of the ground state, one at $T_c = T_0$ and another with a higher rate at $T_B < T_0$, suggesting something like a {\it two-step} BEC due to the spatial anisotropy of the system \cite{Ketterle,Kurt}.
%
% which suggests a {\it second condensation} or a {\it two step-condensation} \cite{Ketterle}. 
%
%
%
Additionally, for $T_B > T_0$ the BEC critical temperature $T_c > T_0$, while for $T_B >> T_0$ the behavior of the CF approaches that of the CF for a constant gap.

In Fig. \ref{fig:FcGapDB} we plot the CF using the $\Delta_{DB}(T)$ gap for several $T_D$ values. For $T_D >> T_0$ the CF tends to that of a constant gap, while for small values of $T_D$ it tends to the case without gap. We note that the discontinuity observed at $T_B$ in the derivative for the ungapped BCS case,  
%in the value of the temperature at which the gap becomes zero, that is, when the fraction of the condensate joins the curve without gap, for the present case
 transforms only into a smooth change of concavity, which avoids touching the CF of the zero gap case. 
  %And never joins the curve with damped gap to the curve without gap since it was chosen that the gap becomes zero at $T_B=10T_0$.
   However, for sufficiently small $T_D$ the damped-gap curves closely approximate the zero gap curve immediately after the concavity changes, as shown by curves $T_D=0.25T_0$, $0.35T_0$ and $0.50T_0$.
From Fig. \ref{fig:FcGapDB} we notice that for all the curves the change in concavity shows up at a temperature greater than $T_D$ but less than $T_c$. Using the damped BCS gap we note that for $T_D$ larger than $1.71T_0$ we don't observe a change in concavity as 
 $T_D \geq T_c$ (see Fig. \ref{TcvsTdD1D2D3Gap}).
 %, i.e., a change of concavity will  be observed for $T_D<T \leq T_c$.
 %, then, for these temperatures there is no change of concavity. 

%  POR QUE DESAPARECE EL CAMBIO DE CONCAVIDAD?(RESPUESTA: EL CAMBIO DE CONCAVIDAD OCURRE POR ARRIBA DE TD. EN TD=1.7 LA TEMPERATURA DE AMORTIGUAMIENTO ALCANZA EL VALOR DE TC QUE ES 1.70579. PARA TD=1.8 O MAYORES TD ES MAYOR QUE TC)
%
\begin{figure}[htb]
	\vspace{-1.8cm}
	\centerline{\epsfig{file=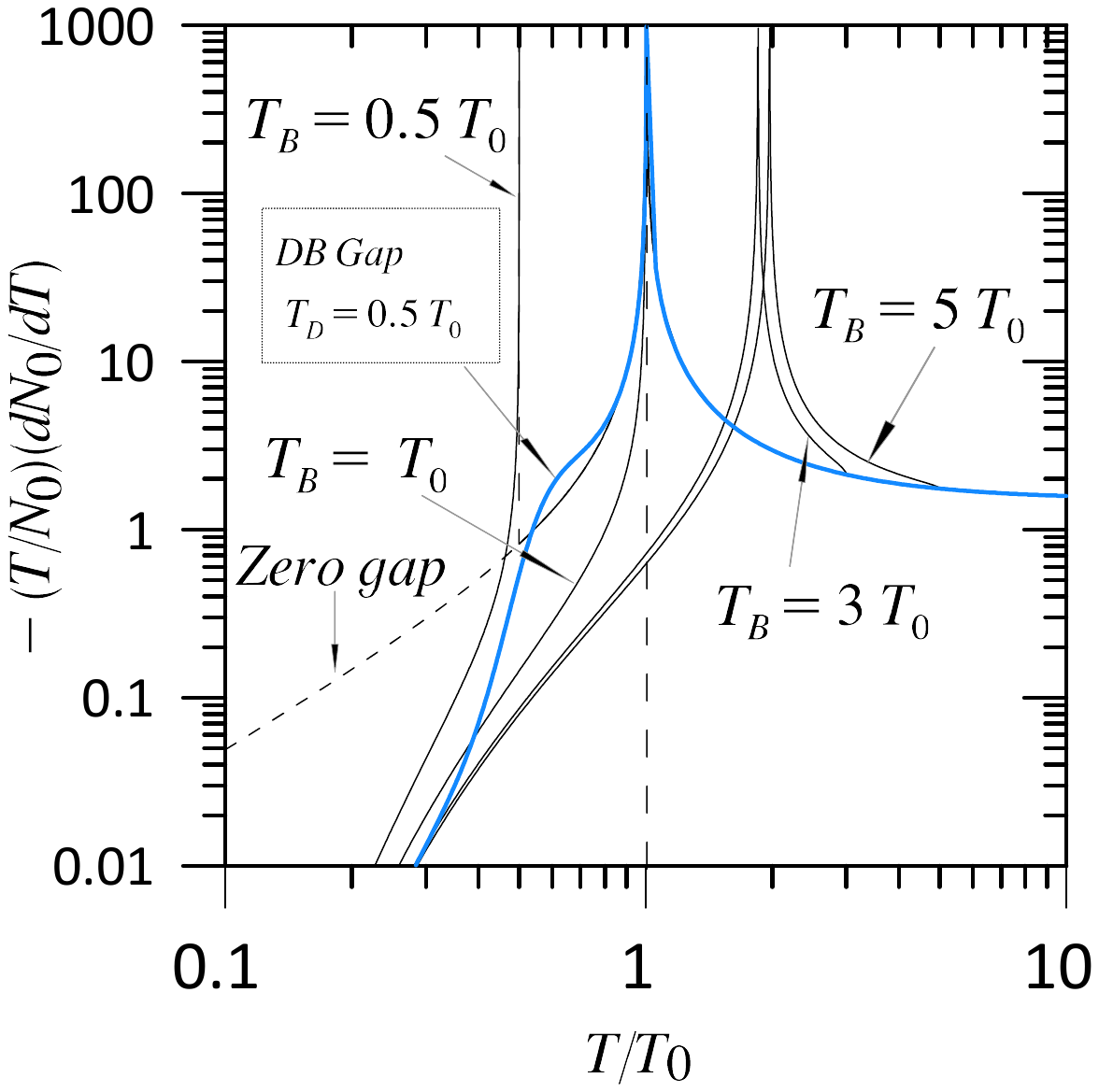,height=3.8in,width=3.7in}}
	\vspace{-0.7cm}
	\caption{Ratio $-(T/N_0)dN_0(T)/dT$ using the zero gap (short dashed line); the undamped BCS gap for values of $T_B/T_0=$ 0.5, 1, 3 and 5; and the damped BCS gap for $T_D = 0.5T_0$ (blue line).}
	\label{fig:dN0}
\end{figure}

Finally, in order to better visualize the divergences in the temperature derivative of the CFs, in Fig. \ref{fig:dN0} we plot the dimensionless ratio $-(T/N_0)dN_0(T)/dT$ for a Bose gas with zero gap 
%and an undamped BCS gap, which 
%. For both cases the ratios 
which diverges at the BEC critical temperature $T_0$ while for the undamped BCS gaps with $T_B \leq T_0$ the ratio has two singularities: at $T=T_B$ and at $T_0$ but, in both cases, for $T>T_B$ the ratio returns to the values of the zero gap ones, as is shown for $T_B=0.5T_0$. When $T_B > T_0$ the ratio presents only one infinity at its BEC critical temperature $T_c > T_0$.  In the same Fig. \ref{fig:dN0} we plot a curve for the ratio using a damped BCS gap with $T_D = 0.5T_0$ to show that the infinity observed at $T_B$ with the BCS gap, transforms into a simple inflection point at $T_D$. 

%\bigskip

\subsection{Gap effect on the chemical potential}

\begin{figure}[htb]
	\vspace{-2.1cm}
	\centerline{\epsfig{file=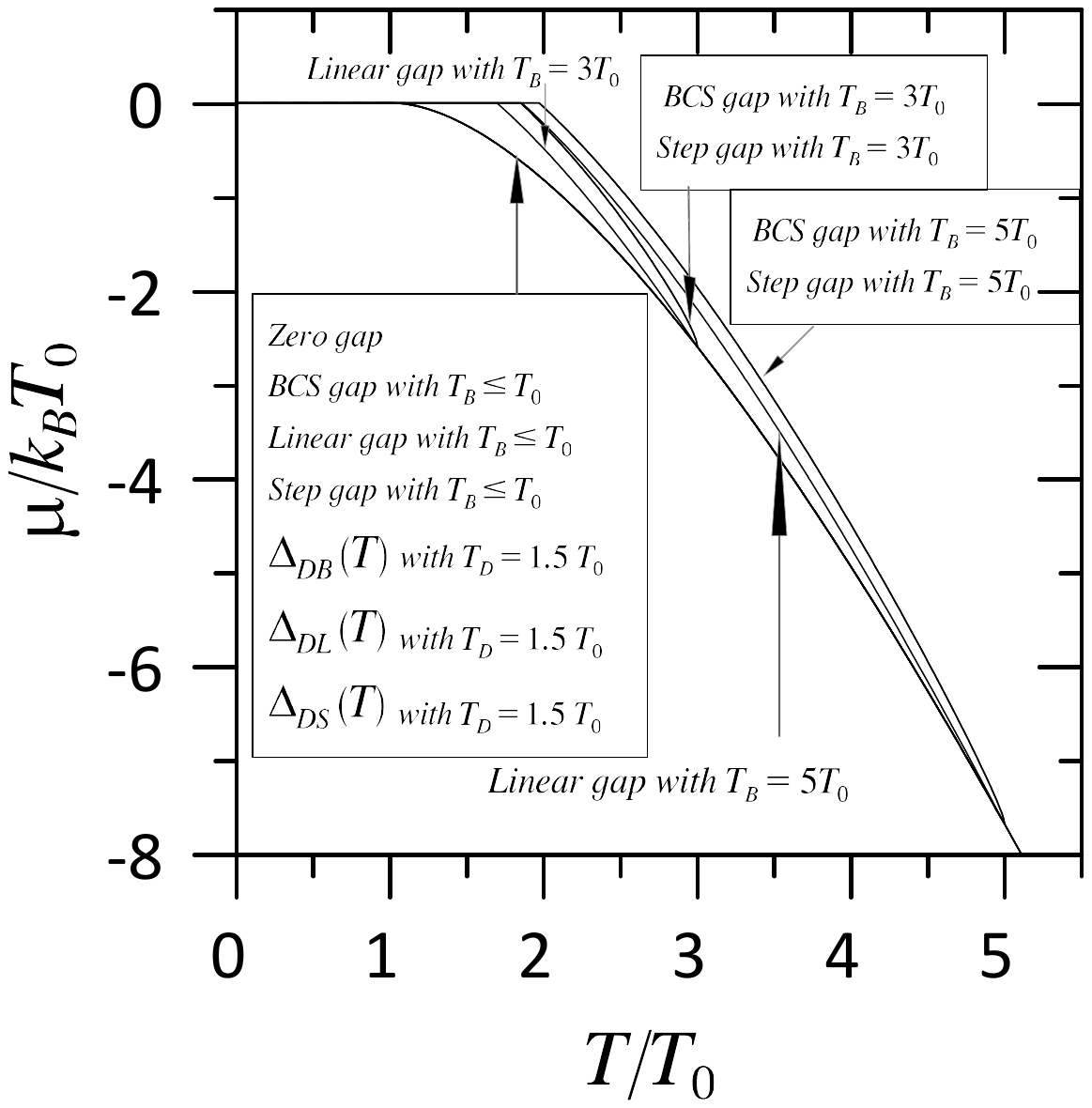,height=3.8in,width=3.8in}}
	\vspace{-0.7cm}
	\caption{Chemical potential as a function of temperature for BCS, step and linear gaps with different values of $T_B$ while for the  $\Delta_{DB}(T)$, $\Delta_{DL}(T)$ and $\Delta_{DS}(T)$ gaps, $T_B = 10 T_0$.}
	\label{fig:MuSeisGaps}
\end{figure}
In Fig. \ref{fig:MuSeisGaps} we show the gap effect on the behavior of the chemical potentials for each case as functions of temperature, for all six damped and undamped gaps.
 %the BCS, step, linear, $\Delta_{DB}(T)$, $\Delta_{DL}(T)$, and $\Delta_{DS}(T)$ gaps. 
 For the Bose gas, when $T_B = T_0$, the chemical potential is the same using any of the undamped BCS, Step, and Linear gaps or using   
  %we substitute $T_B = T_0$ the chemical potential has the same behavior with these 
  %as well as 
  any of the damped $\Delta_{DB}(T)$, $\Delta_{DL}(T)$, and $\Delta_{DS}(T)$ gaps, with $T_D = 1.5 \, T_0$. However, using any of the three undamped gaps, when we substitute $T_B =3 \, T_0$ or $5 \, T_0$ the chemical potential curves merge to the zero-gap one only for temperatures $T \geq T_B$; at $T_B$ 
 %  at  after $T_B$ all the cases are put together again. 
the temperature derivative of the chemical potential diverges. 
% At such a junction the slope of the curves is infinite.???
%
\begin{figure}[htb]
	\vspace{-1.8cm}
	\centerline{\epsfig{file=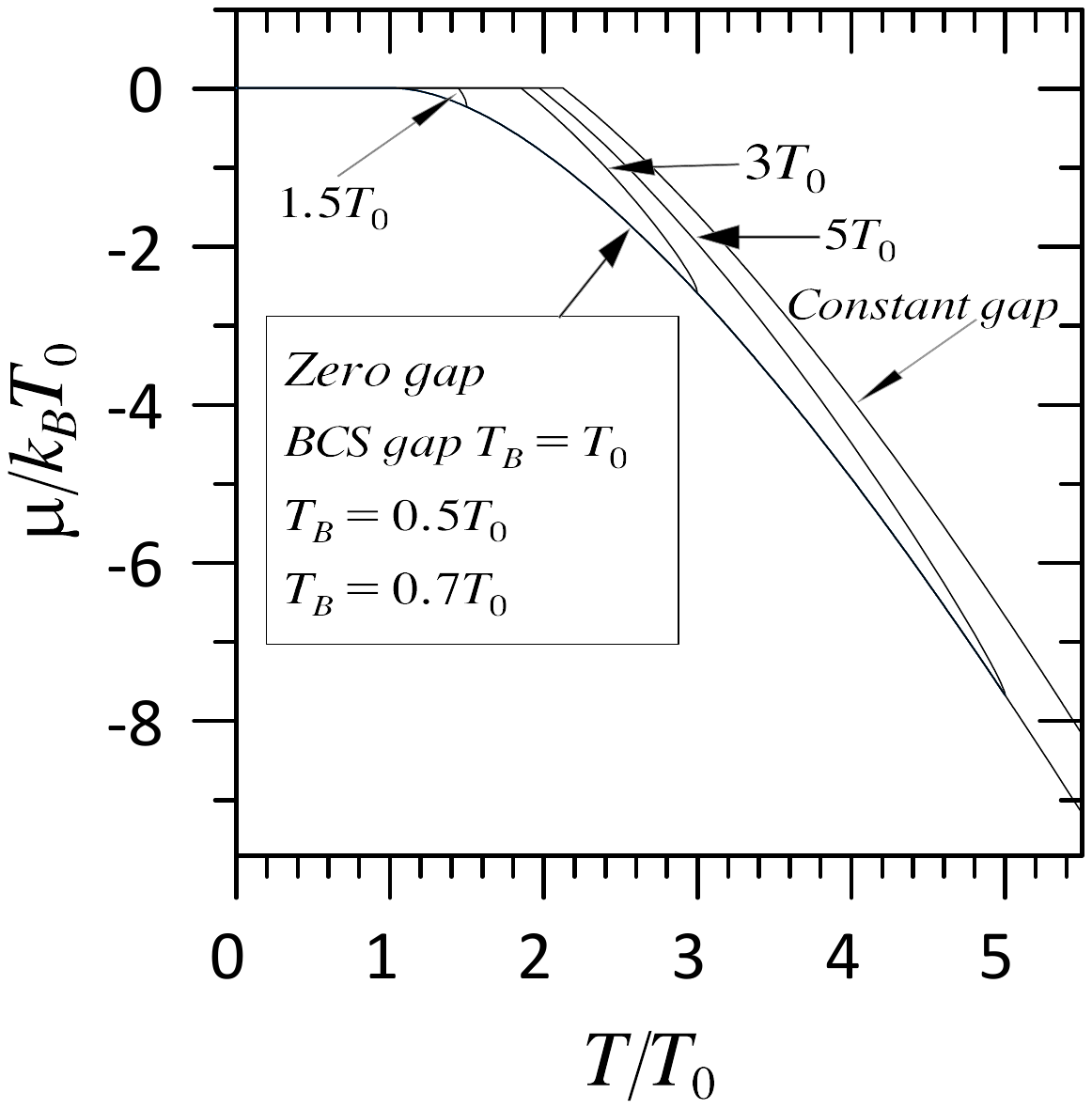,height=3.8in,width=3.8in}}
	\vspace{-0.7cm}
	\caption{Chemical potential as a function of temperature for BCS gap for several values of $T_B$ compared to the zero and constant gap cases.}
	\label{fig:MuGapBCS}
\end{figure}
\begin{figure}[htb]
	\vspace{-1.4cm}
	\centerline{\epsfig{file=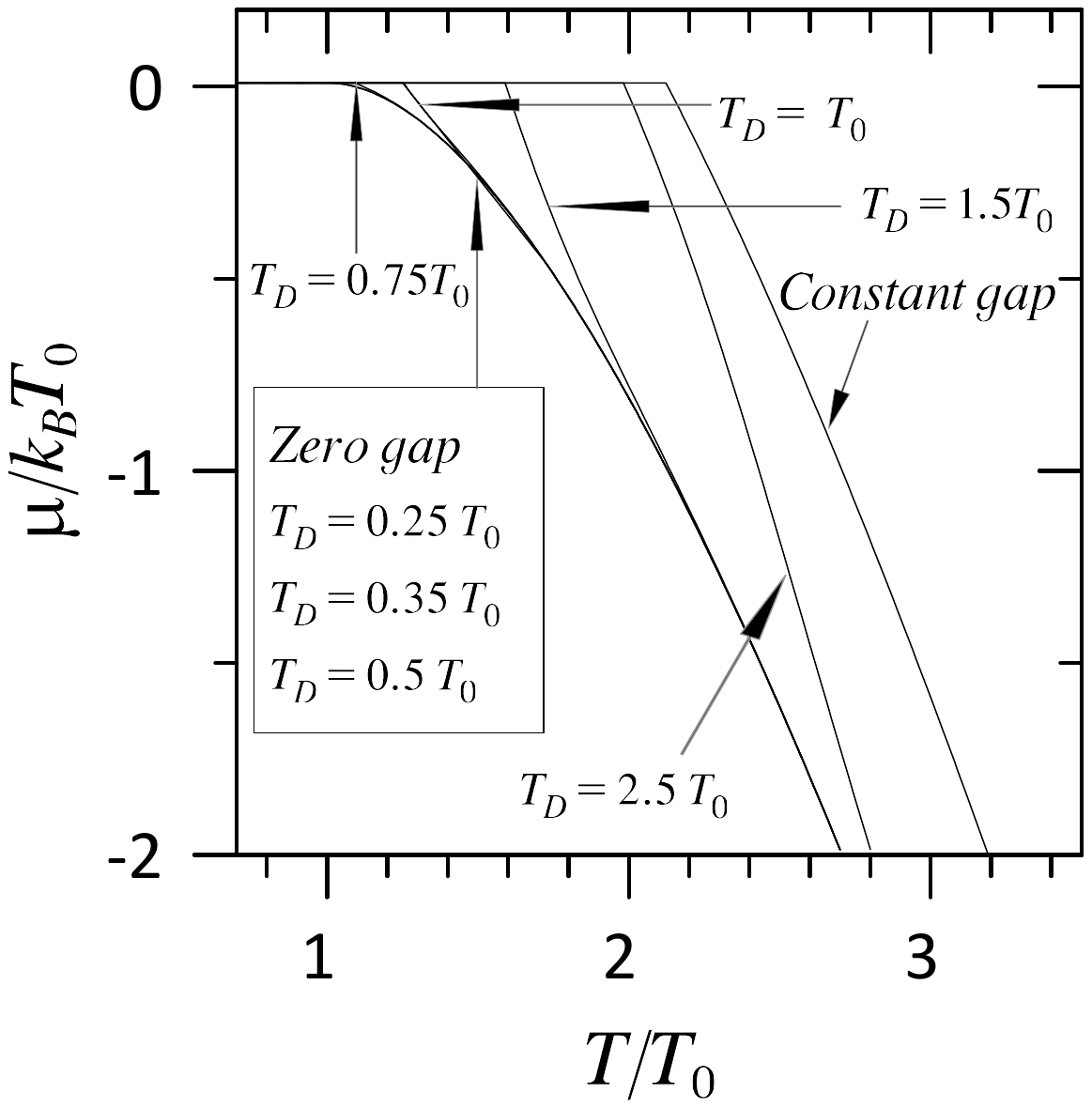,height=3.8in,width=3.8in}}
	\vspace{-1.4cm}
	\caption{Chemical potential as a function of temperature for $\Delta_{DB}(T)$ gap for several values of $T_D$ compared to the zero and  constant gap cases.}
	\label{fig:MuGapDB}
\end{figure}
In Figs. \ref{fig:MuGapBCS} and \ref{fig:MuGapDB} we show the behavior of the chemical potential as a function of temperature when we use the BCS gap varying the value of $T_B$, as well as the $\Delta_{DB}(T)$ gap varying the damping temperature $T_D$.  For the BCS gap case if $T_B \leq T_0$ the chemical potential reduces to the gapless case. We do the same for the $\Delta_{DB}(T)$ gap case, but now for $T_D$ smaller than $\sim 0.5T_0$. If $T_B$ or $T_D$ take larger values, the chemical potential approaches the constant gap case. Again, the main difference between the BCS and $\Delta_{DB}(T)$ cases is how smoothly their chemical potential curves merge with that for the zero gap.
% chemical potential different specific heat curves come together.
%the main difference between the BCS and $\Delta_{DB}(T)$ cases is the smoothness in the change of inclination of the curves.   ????

\subsection{Gap effect on the internal energy}
\begin{figure}[htb]
	\vspace{-1.08cm}
	\centerline{\epsfig{file=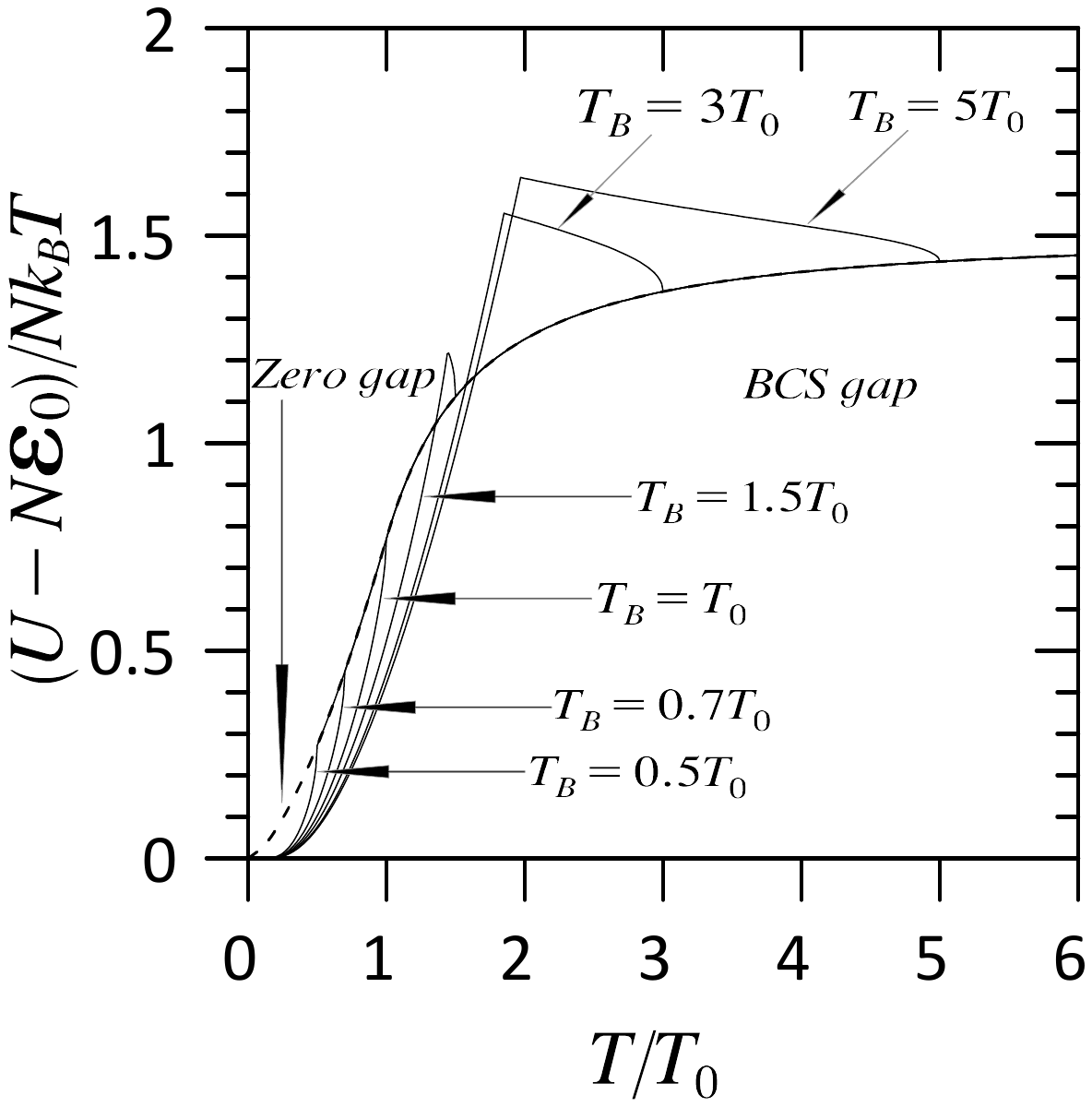,height=3.6in,width=3.60in}}
	\vspace{-1.3cm}
	\caption{Internal energy divided by $N k_B T$ as a function of temperature for the BCS gap with $T_B = 0.5 \, T_0, 0.7 \, T_0, T_0, 1.5 \, T_0, 3\, T_0$, and $5 \, T_0$.  }
	\label{fig:UMenosEceroNBCS}
\end{figure}

\begin{figure}[htb]
	\centerline{\epsfig{file=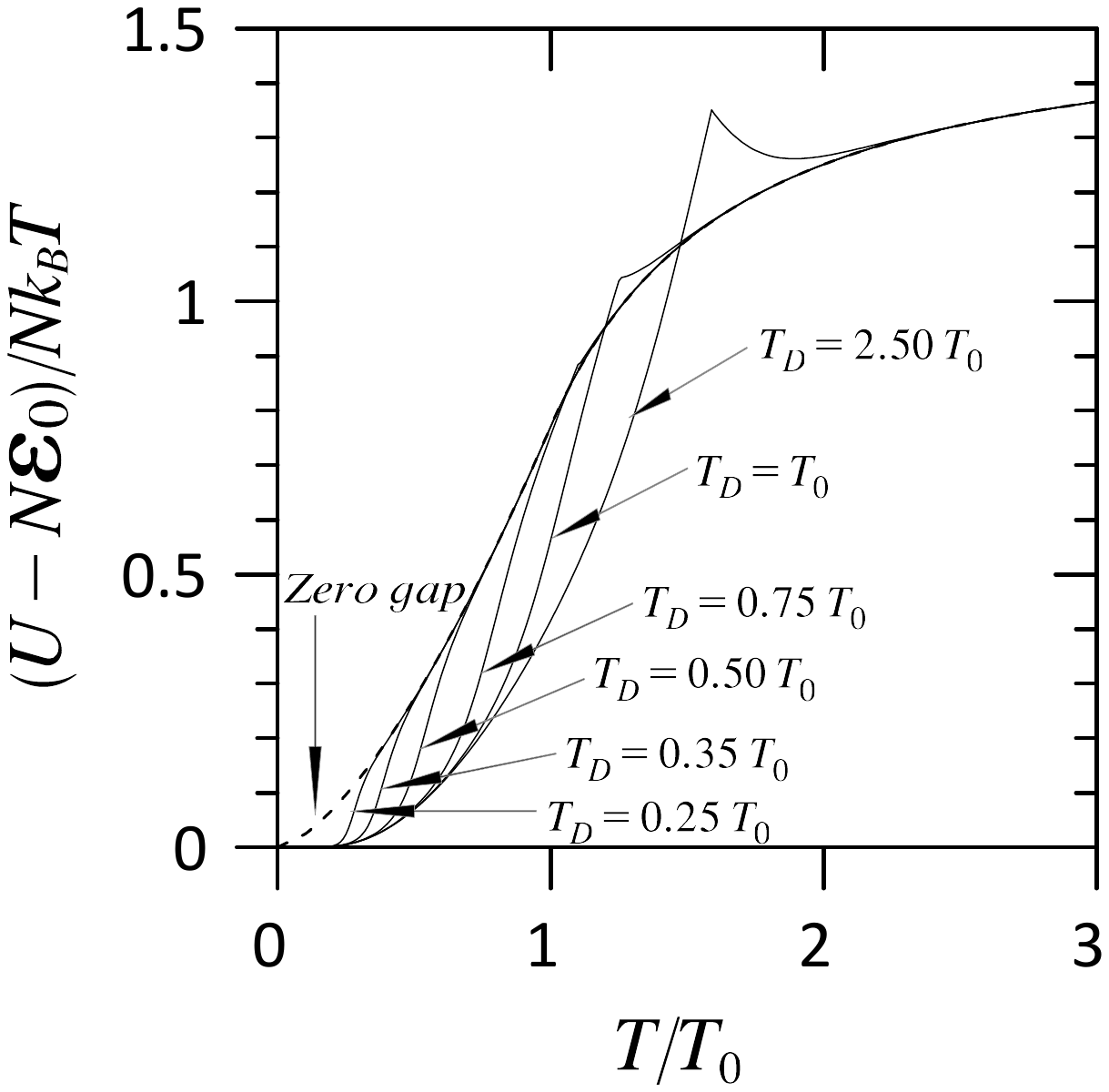,height=2.8in,width=4.5in}}
	\vspace{-0.3cm}
	\caption{Internal energy divided by $N k_B T$ as a function of temperature using $\Delta_{DB}(T)$ gap with $T_D = 0.25 \, T_0, 0.35 \, T_0, 0.50 \, T_0, 0.75 \, T_0, T_0$, and $2.50 \, T_0$.}
	\label{fig:UMenosEceroNDeltaDB}
\end{figure}

\begin{figure}[htb]
	\vspace{-2.05cm}
	\centerline{\epsfig{file=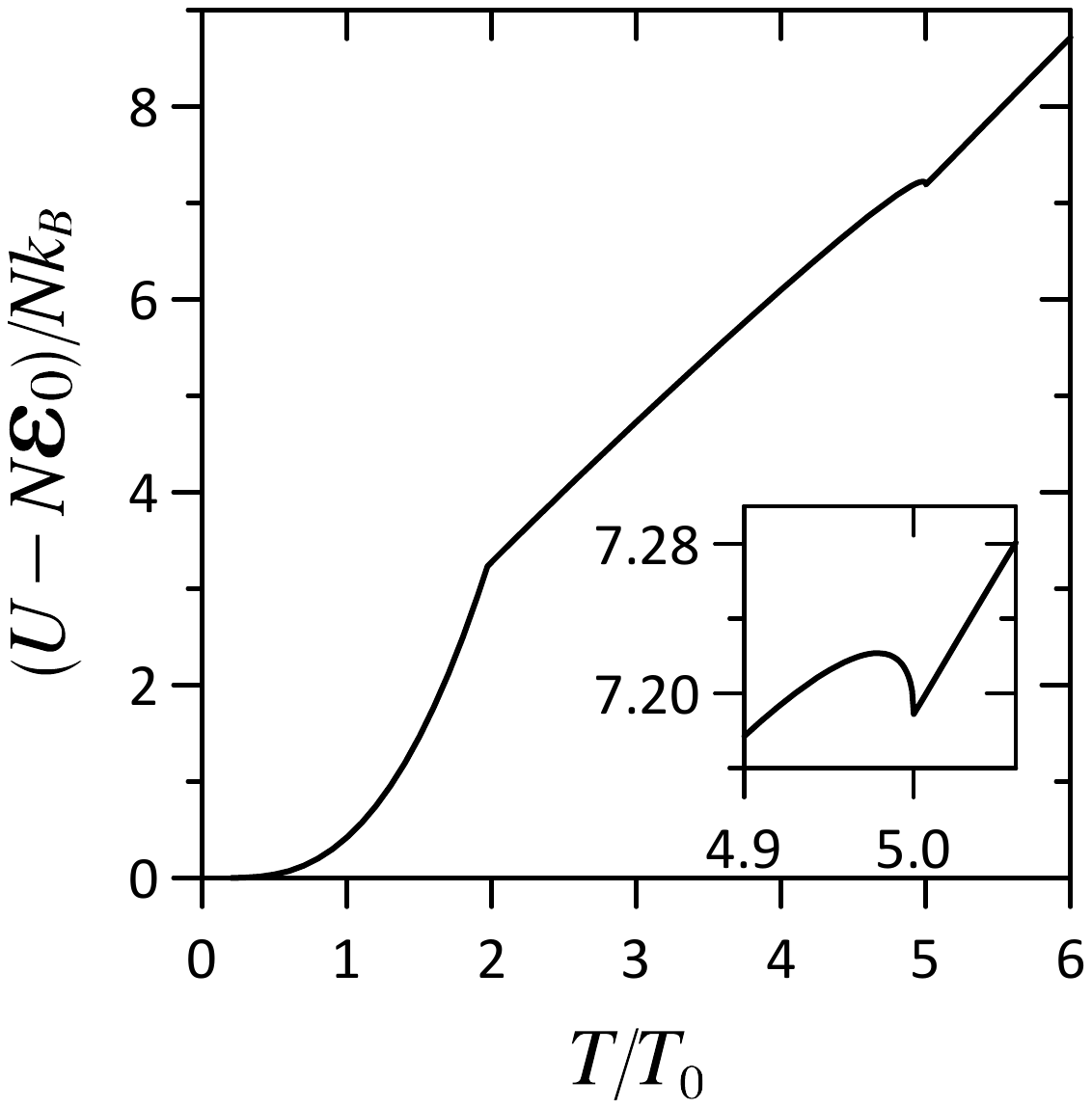,height=3.8in,width=3.6in}}
	\vspace{-0.7cm}
	\caption{Internal energy using BCS gap with $T_{B}=5T_0$. The inset shows a negative infinite slope at $T=T_{B}$. }
	\label{fig:UtConGapBCSTBCS5}
\end{figure}

Figures \ref{fig:UMenosEceroNBCS} and \ref{fig:UMenosEceroNDeltaDB} show the internal energies divided by $N k_B T$ and referred to the ground state energy $N \varepsilon_0$, as functions of temperature for the undamped and damped BCS gaps, respectively. For every case, the internal energy divided by $k_B T$ tends to $3/2$ at large temperatures, which is consistent with the principle of equipartition of energy. In Fig. \ref{fig:UMenosEceroNBCS}, we show the internal energy divided by $k_B T$ using the undamped BCS gap for various values of $T_B$ 
at which there is a discontinuity in its derivative as it merges to the zero gap curve, i.e. the derivative is positive infinite for $T_B < T_0$ and negative infinite for $T_B > T_0$.  In Fig. \ref{fig:UtConGapBCSTBCS5} we plot the internal energy, without dividing it by $k_B T$,  
to better visualize its behavior as well as that of its derivative; the inset is a close up of the internal energy curve near $T_B$.
This strange behavior of the internal energy derivative at $T_B$ is not observed using the damped $\Delta_{DB}(T)$ gaps since all curves meet smoothly with the zero gap curve,  Fig. \ref{fig:UMenosEceroNDeltaDB}. 
We also note that using the undamped gap, for $T_B > T_0$ the internal energy per particle divided by $k_B T$ shows a peak at the corresponding transition temperature from which the internal energy increases more slowly than $k_B T$,  even though at higher temperatures it grows again. This peak is also observed using the $\Delta_{DB}(T)$ gap, but it gradually disappears as $T_D$ decreases.

\subsection{Gap effect on the calculation of the specific heat}
\begin{figure}[htb]
	\vspace{-1.9cm}
	\centerline{\epsfig{file=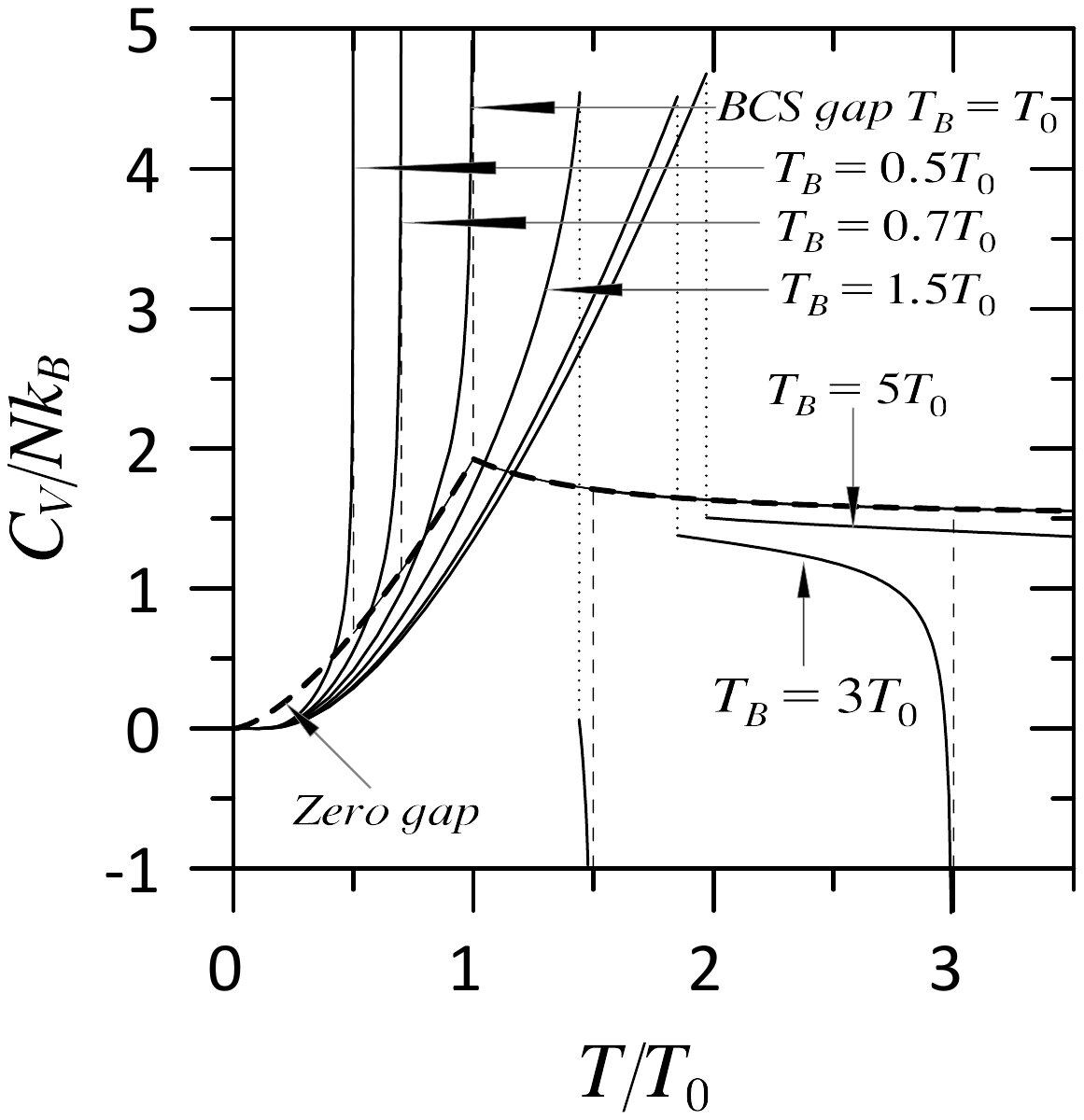,height=3.8in,width=3.6in}}
	\vspace{-0.5cm}
	\caption{Isochoric specific heat using the BCS gap forh various values of $T_{B}$. As a reference, we also show the specific heat calculated with the zero gap (thick dashed line). Dot lines indicate a jump and thin dashed lines indicate a divergence in the specific heat.}
	\label{fig:CvGapBCStB05y07y1y1.5y3y5}
\end{figure}
To continue exploring the behavior of thermodynamic properties when the gap has the shape proposed by BCS, we plot the specific heat versus temperature in Fig. \ref{fig:CvGapBCStB05y07y1y1.5y3y5} for various values of $T_B$. For every $T_B < T_0$, the BEC critical temperature is $T_0$ where the specific heat shows a peak in addition to a positive singularity at $T_B$.
For $T_B=T_0$ the specific heat shows only a positive singularity with a critical exponent equal to $\alpha = 1/2$, which is inherited from the divergence of the temperature derivative of the BCS gap at $T_B$.
%From this figure we can say that, 
For $T_B > T_0$, the specific heat presents a jump at the corresponding transition temperature plus a negative singularity at $T=T_B$, which comes from the negative infinite slope of the BCS gap at that temperature. 
 %at that temperature, although it shows a peak when going through $T_0$. 
%
 %Another important thing to note is that the negative singularity that each curve has at $T=T_B$ when $T_B \geq T_0$, becomes positive for $T_B < T_0$. And right at $T_B=T_0$ the negative divergence becomes positive and appears at the transition temperature.
 
 % Infinities at the temperature where the gap becomes zero can be avoided by smoothing the gap drop, as is the case with gap $\Delta_{DB}(T)$.

\begin{figure}[htb]
	\vspace{-2cm}
	\centerline{\epsfig{file=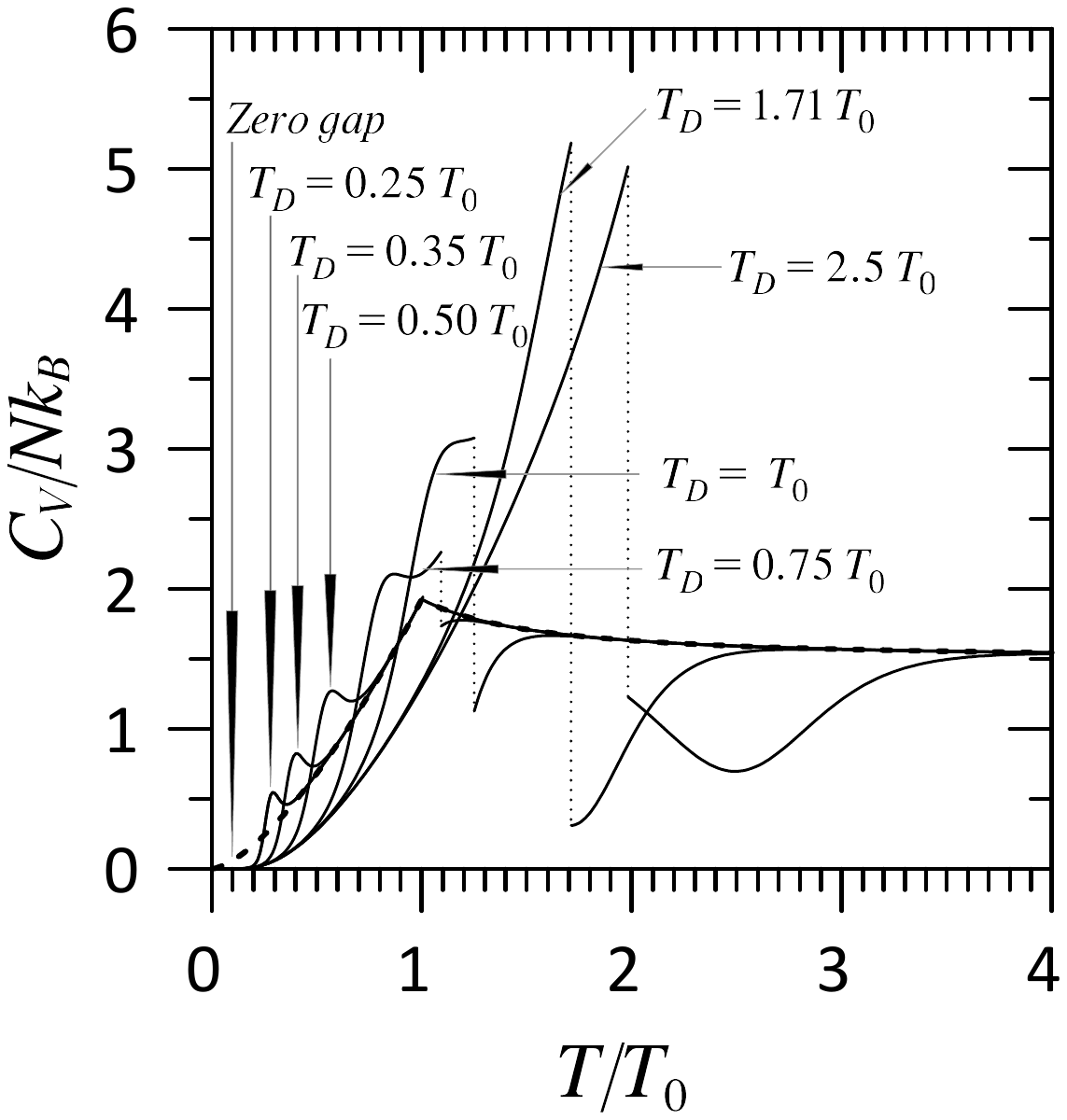,height=3.8in,width=3.6in}}
	\vspace{-0.5cm}
	\caption{Isochoric specific heat using the $\Delta_{DB}(T)$ gap forh several values of $T_D$, besides the gapless case (dashed line). For smaller values of $T_D \simeq 0.5 T_0$  the jump in the specific heat disappears becoming a peak and leaving an additional hump.} 
		 \label{fig:CvGapDB}
\end{figure}

To avoid singularities we use the $\Delta_{DB}(T)$ gap instead of the undamped BCS gap, so the 
 %in the calculation of the specific heat we obtain Fig. \ref{fig:CvGapDB}. 
 infinities observed in Fig. \ref{fig:CvGapBCStB05y07y1y1.5y3y5}  become 
 %that we had when using the BCS gap are now 
 maximum and minimum values as observed in Fig. \ref{fig:CvGapDB}. 
 
 For critical temperature values greater than $T_0$ the specific heat has a jump which increases as the critical temperature $T_c$ raises until it reaches a maximum at $T_D/T_0 = T_c/T_0 =  1.71$. For every $T_D \lesssim 0.5 T_0$, $T_c =T_0$ and the specific heat has a hump at $T_D$ after which it merges to the $C_V$ curve of the  zero gap case.    
 
% Para valores de temperatura crítica mayores que T_0 (TD/T0 aprox 0.5) el calor específico presenta un salto que aumenta conforme aumenta la temperatura crítica hasta alcanzar un máximo donde TD/T0 = Tc/T0 = TD/T0 =  1.71. k
% Para todos los valores de TD <apr 0.5 la Tc = T0 and El Cv muestra un hump at TD para luego juntarse con aquella curva del Cv del  . 

We note that every $C_V$ curve presents an exponential behavior for temperatures near zero, while for much higher temperatures they tend to the 3/2 value as expected in the classical limit.  
 
% More specifically, there is a minimum value near the damping temperature if $T_D > 0.5T_0$, but if $T_D \leq 0.5T_0$ it becomes a maximum. When $T_D > 0.5 T_0$ appears a jump in the specific heat at the critical temperature $T_c$. And for smaller values of $T_D \simeq 0.5T_0$ the jump becomes a peak at $T_c$. 
 %However, there is an extremely small jump in the specific heat at $T_c$ because at these temperatures the fraction of condensate with damped gap is too close to the fraction of condensate without gap but they don't join (see Fig. \ref{fig:FcGapDB}). 

%And finally, in $T_D = T_0$ the maximum and the jump coincide in $T_c$.

We observe that although the cases with zero gap and BCS gap with $T_B \leq T_0$ have the same chemical potential, the behavior of their corresponding $C_V$ is very different, i.e., 
the energy gap has a strong influence on the specific heat behavior, but not on the chemical potential. 

%When the gap is zero, the transition to the condenseate state is of first order and at $T_c$ the specific heat show a discontinuity in its derivative
%\cite{Aguilera-Navarro}.
% Making $\Delta=0$ in Eqs. (\ref{cvbajotc}) and (\ref{cvarribatc}) we reproduce this discontinuity in the derivative of the specific heat, see Fig. \ref{fig:Cvgapcero}. 

%\begin{figure}[Htb]
%	\centering
%	\begin{minipage}[b]{0.45\linewidth}
%		\centering
%		\includegraphics[width=\textwidth]{Cv con gap cero.pdf}
%		\caption{ Specific heat when the gap of energy is zero.}\label{fig:Cvgapcero}
%	\end{minipage}
%	\quad
%	\begin{minipage}[b]{0.45\linewidth}
%		
%		\includegraphics[width=\textwidth]{mu con gap cero.pdf}
%		\caption{Chemical potential with zero energy gap.}\label{fig:mu con gap cero}
%	\end{minipage}  
%	
%\end{figure}

%When the gap is different from zero and constant, the transition is of second order and the specific heat has a discontinuity at the critical temperature \cite{Martinez-Herrera}. 
\begin{figure}[htb]
\vspace{-1.8cm}
\hspace{-1.0cm}	\centerline{\epsfig{file=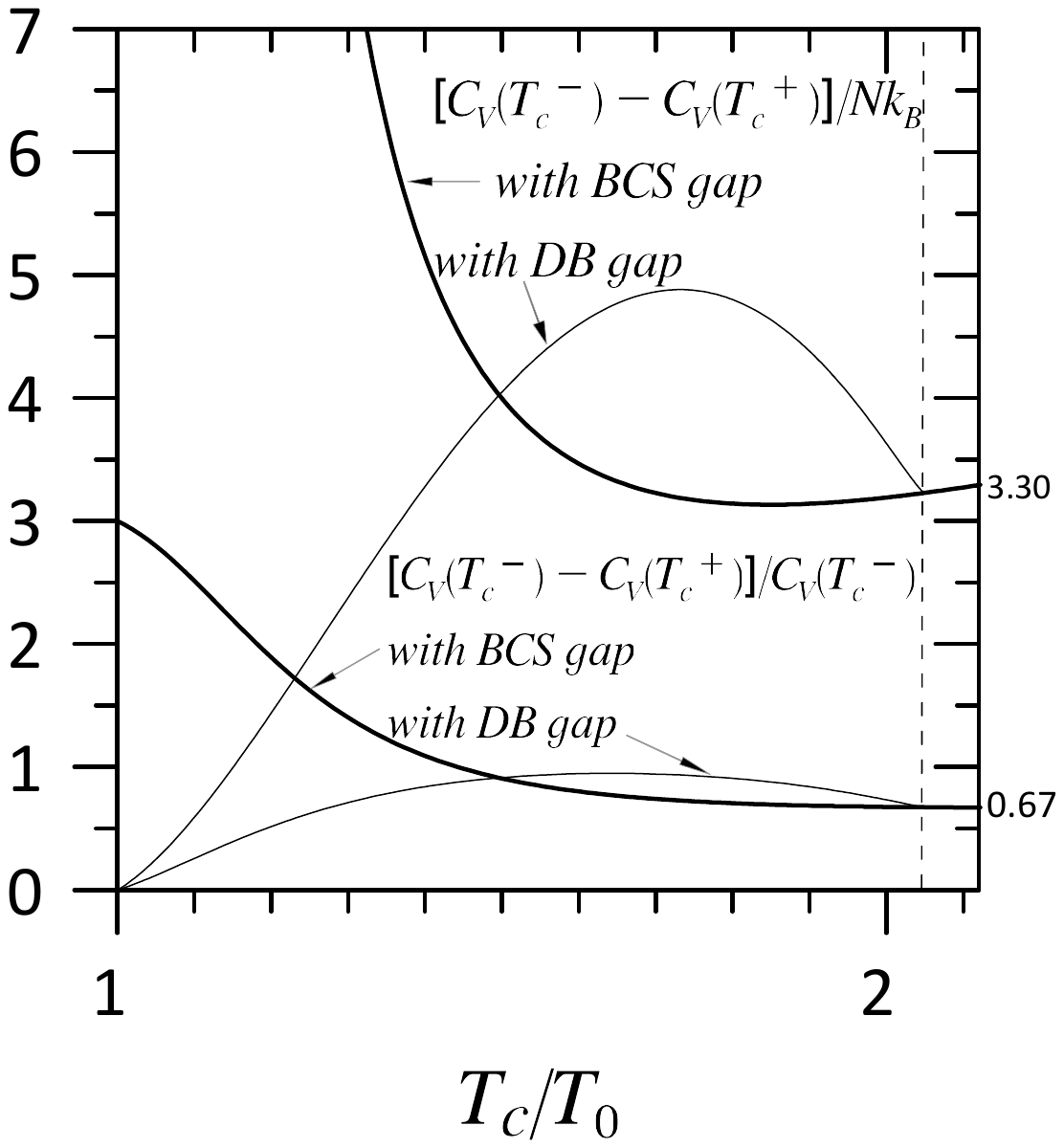,height=3.8in,width=3.8in}}
\vspace{-0.5cm}
	\caption{Isochoric heat jump magnitude at the BEC critical temperature $Tc/T_0$ as well as its magnitude divided by $C_V(T_c)$, both as functions of $T_c/T_0$, for the BCS  (thick line) and DB gap (thin line) cases.
	}
	\label{fig:SaltoDeCvVsTc}
\end{figure}

Finally, Fig. \ref{fig:SaltoDeCvVsTc} shows the specific heat jump and its magnitude divided by $C_V(T^{-}_c)$, both as functions of $T_c/T_0$, which in turn is a function of $T_B$ for the undamped BCS gap and of $T_D$ for the DB gap cases. As we increase the values of $T_B$ and/or $T_D$, the BCS and DB gaps tend to a constant gap $\Delta_0$ that corresponds to a BEC critical temperature of 2.12 $T_0$ where the values $[C_V(T^{-}_c)-C_V(T^{+}_c)]/Nk_B = 3.30$ and $[C_V(T^{-}_c)-C_V(T^{+}_c)]/C_V(T^{-}_c) = 0.67$.
%
%discontinuity passing through the transition temperature as a function of the same $T_c$. 
%In Fig. \ref{fig:SaltoDeCvVsTc} are shown 
%In the BCS gap case we vary the value of the temperature where the gap dies, $T_B$, and for the DB gap case we vary the value of the damping temperature $T_D$. With this we achieve a variant $T_c$.
%
% From Figs. \ref{fig:TcvsTBConstantBCSAndLinearGap} and \ref{TcvsTdD1D2D3Gap} we can see that, by increasing $T_B$ or $T_D$, the critical temperature tends to $T_c = 2.12T_0$, which is the value of the critical temperature for the constant gap case. And the jump of $C_V$ at this temperature is $[C_V(T^{-}_c)-C_V(T^{+}_c)]/Nk_B = 3.24$, which is the value for the constant gap case.
Note that  for every $T_c \leq T_0$, the specific heat shows a zero jump using the damped BCS gap, while using the undamped BCS gap it presents an infinite jump. 
In adition to this, when using the undamped BCS gap the specific heat jump as a function of the BEC critical temperature, decreases until it reaches a minimum at 
 $T_c/T_0 = 1.85$, for $T_B = 3 \, T_0$. While using the damped BCS gap, with $T_D = 1.75 \, T_0$ and $T_B = 10 \, T_0$, the jump increases from $T_c = T_0$ until it reaches $T_c = 1.73 \, T_0$ where the specific heat jump shows a maximum from where it decreases to reach the value of 3.23 at  $T_c = 2.05 \, T_0$, which is the maximum critical temperature for $T_B =  10 \, T_0$.  

%It is worth mentioning that in Ref. \cite{Kim Faeth and Stewart} it is mentioned that the jump in the specific heat grows with $T_c$ for Ba(Fe$_1-x$Co$_x$)$_2$As$_2$.

%Keeping the energy gap constant in the two equations for specific heat, we reproduce this transition in Fig. \ref{fig:Cvgapconstante}, where $\Delta=k_BT_0$. 

%\begin{figure}[Htb]
%	\centering
%	\begin{minipage}[b]{0.45\linewidth}
%		\centering
%		\includegraphics[width=\textwidth]{Cv con gap constante.pdf}
%		\caption{ Specific heat when the gap of energy is constant $\Delta= k_B T_0$.}\label{fig:Cvgapconstante}
%	\end{minipage}
%	\quad
%	\begin{minipage}[b]{0.45\linewidth}
%		
%		\includegraphics[width=\textwidth]{mu con gap constante.pdf}
%		\caption{Chemical potential when the gap of energy is constant $\Delta= k_B T_0$.}\label{fig:mu con gap constante}
%	\end{minipage}  
%	
%\end{figure}
%
%\bigskip
%\pagebreak
\section{Conclusions}
We have given the expressions for 
%After calculating 
the BEC critical temperature, the condensate fraction, the chemical potential, the internal energy and the isochoric specific heat for a 3D Bose gas whose particles have the energy momentum relation $\varepsilon(k) =\varepsilon_0+\Delta (T)+\hbar^2 k^2/2m$, i.e., a temperature dependent gap between the ground and the first excited state energies. These expressions are used to calculate the mentioned thermodynamic properties for six gaps: the BCS, Linear and Step gaps, each one in its undamped and damped versions.

To begin with, we confirm that for a temperature independent constant gap: a) All the calculated thermodynamic properties are constant ground state energy $\varepsilon_0$ independent, except for the internal energy, which is measured from the reference energy, $N \varepsilon_0$, instead of from the zero ground state energy of an IBG, b) A constant gap notoriously increases the magnitude of the BEC critical temperature, where the isochoric specific heat shows a jump, while for $T$ near zero  its temperature dependence is exponential rather than proportional to $T^{3/2}$,  as for a 3D ideal Bose gas, as already observed in Ref. \cite{Martinez-Herrera}

%	\item Below the critical temperature and near to zero absolute, the temperature dependence of the specific heat is exponential with an exponent proportional to the gap, however, when the gap becomes zero the temperature dependence of the specific heat becomes that of the IBG, i. e., the specific heat behaves as a power of temperature. 

%\noindent {\bf The decay with infinite slope.} 
The decay of the undamped BCS gap with infinite slope at $T_B$ causes abrupt changes in the behavior of the thermodynamic properties of the boson gas as they go through the temperature $T_B$. 
In order to explore the effect of Cooper pairs creation before they reach the BEC critical density we propose a smooth transition from a non-zero gap to an equal to zero one at $T_B$, multiplying the undamped BCS gap by an exponential decreasing function of temperature.
 %for temperatures larger than $T_c$. 
 In this way we prevent the appearence of unwanted infinities at $T_B$ in the thermodynamic properties of the Bose gas. 
 %
 %{\bf Three undamped gaps vanishing at $T_B \leq T_0$.}   
 
 For the three types of undamped gaps studied which vanish at $T_B \leq T_0$, the  BEC critical temperature $T_c$ is equal to $T_0$ and gap shape independent.
 %In other words, if the gap vanishes before $T_0$,  the amount of energy $k_BT_0$ needed to change is not present ??? OJO the IBG does not know enough about the gap to change its BEC critical temperature. 
 However, the gap has an important effect on the behavior of the condensate fraction (CF) which, for temperatures $T_B < T < T_0$, is equal to that of the IBG, while for $0 < T < T_B< T_0$ the rate of accumulation of particles in the ground state shows a sharp increase at $T_B$, suggesting something like a {\it two-step} condensation due to the spatial anisotropy of the system. Also at $T_B$ the internal energy shows a peak which turns into a positive infinite divergence in the specific heat. For every $T_B < T_0$, the BEC critical temperature is $T_0$ where the specific heat shows a peak in addition to a positive singularity at $T_B$.
 For $T_B=T_0$ the specific heat shows only a positive singularity with a critical exponent equal to $\alpha = 1/2$, which is inherited from the divergence of the temperature derivative of the BCS gap at $T_B$.  
 % WHAT ABOUT THE OTHERS PROPERTIES?    
 %
 %	{\bf Three damped gaps vanashing at $T_B$.}
%
%\noindent  {\bf Three damped gaps vanishing at $T_D< T_0 < T_B$.}    
%
%        increase in its growth greater than that of the ideal Bose gas.  of bosonsa new no longer exhibits some like a {\it second condensation}.    
%INDICAR LO QUE PASA EN T_B Y EN T_c PARA EL Cv ... \\
%	we soften the behavior of the bosonic system when passing from the region with gap to the region without gap.
%	explore the effect of Cooper pairs creation before to reach the critical density for BEC
%{\bf Three damped gaps vanishing at $T_0 < T_D < T_B$.} 

The damped gaps become zero at $T_B$ which is larger than both $T_0$ and $T_D$ where the gap magnitude is $\Delta_0/2$. For every $T_D$ the critical temperature  $T_c \geq T_0$. In addition, the critical temperature as a function of $T_D$ is larger than $T_D$  untill $T_D$ equals $T_c$, after which $T_c$ becomes smaller than $T_D$. The values of $T_D$ for which $T_D = T_c$ depend on the damped gap shape, as shown in the inset of Fig. \ref{TcvsTdD1D2D3Gap}. On the other hand, for $T_D \lesssim 0.5 T_0$ the $T_c$ equals $T_0$ 
%where the critical temperature curve smootly reachs $T_0$; for temperatures below $T_D = 0.5 T_0$ 
where the CF growths smoothly and remarkably above that of the ideal Bose gas. 

Finally, by associating the boson energy gap with a temperature dependence, one can explore how the size of the specific heat discontinuity at $T_c$ changes as a function of $T_c/T_0$, allowing us to identify the type of phase transition that the system presents. \\

%\end{itemize}
We thank Dr. P. Salas for her careful reading of the manuscript and her valuable comments.
M.A.S. thanks the partial support from grant UNAM-DGAPA-PAPIIT IN114523.


\begin{thebibliography}{99}
%\bibitem{seudogap}...
\bibitem{Das} S. Das and S. Sur,
%A unified cosmological dark sector from a Bose–Einstein condensate
 Physics of the Dark Universe {\bf 42}, 101331 (2023); S. Das and R.K. Bhaduri, arXiv:1808.10505v3 [gr-qc] Bose-Einstein condensate in cosmology.
\bibitem{Hwang} Hwang, J. %Superconducting coherence length of hole-doped cuprates obtained from electron–boson spectral density function. 
Sci Rep 11, 11668 (2021).
\bibitem{Lee} R. Friedberg and T.D. Lee, Phys. Rev. B {\bf 40}, 6745 (1989); R. Friedberg, T.D. Lee and H.C. Ren, Phys. Lett. A {\bf 152}, 417 (1991).
\bibitem{Tolmachev} V.V. Tolmachev, Phys. Lett. A {\bf 266}, 400 (2000); M. de Llano and V.V. Tolmachev, Physica A {\bf 317}, 546 (2003).
\bibitem{London}F. London, Nature {\bf 141}, 643 (1938).
\bibitem{London2} F. London, {\it Superfluids} Vol. II (John Wiley \& Sons, New York, 1954)p. 53.
\bibitem{japoneses} M. Toda, Prog. Theor. Phys. {\bf 6}, 458 (1951); T. Matsubara,  Prog. Theor. Phys. {\bf 6}, 458 (1951). % japoneses
\bibitem{gapjustification}  M. Girardeau and R. Arnowitt, Phys. Rev. {\bf 113}, 755 (1959).  %many theoretical efforts gap justification.
\bibitem{HP}N. M. Hugenholtz and D. Pines, Phys. Rev. {\bf 116}, 489 (1959).%No encuentran gap.
\bibitem{juan} J. García-Nila, M. Sc. (Physics) Thesis, (Universidad Nacional Autónoma de México, Ciudad de México, 2019).
\bibitem{Pathria} R. K. Pathria and P. D. Beale, {\it Statistical Mechanics} Third Edition  (Elsevier, Oxford, 2011), pp. 664-667.
\bibitem{Martinez-Herrera} J.G. Martínez-Herrera, J. García-Nila and M. A. Solís, Phys. Scr. {\bf 94}, 075002 (2019). 
\bibitem{Aguilera-Navarro} V. C. Aguilera-Navarro, M. de Llano and M. A. Solís, Eur. J. Phys. {\bf 20}, 177 (1999).
\bibitem{seudogap} N. Doiron-Leyraud, et al., Nature Communications {\bf 8}, 2044 (2017). DOI: 10.1038/s41467-017-02122-x, www.nature.com/naturecommunications
\bibitem{Dougherty} R. C. Dougherty, and J. D. Kimel, arXiv-1212.0423-Temperature dependence of the superconductor energy gap.
\bibitem{Ketterle} W.J. Mullin and A.R. Sakhel, J. Low Temp. Phys., {\bf 166} 125 (2012); R. Ramakumar and A.N. Das, Phys. A {\bf390} 208 (2011); K. Shiokawa,  J. Phys. A: Math. Gen. {\bf 33} 487 (2000);
	%. Generalized bose{einstein condensation.
     N.J. van Druten and W. Ketterle, Phys. Rev. Lett. {\bf 79}, 549 (1997).
     % On multistep Bose-Einstein condensation in anisotropic traps.
% J. Phys. A: Math. Gen., 33:487{506. 3, 6, 13.
\bibitem{Kurt} C. Kurt et al., Phys. Scr. {\bf 100} 015289 (2025). 
\bibitem{Kim Faeth and Stewart} J. S. Kim, B. D. Faeth, and G. R. Stewart, Phys. Rev. B {\bf 86}, 054509 (2012). 
%Specific Heat Discontinuity vs Tc in Annealed Ba(Fe$_1-x$Co$_x$)$_2$As$_2$.
\bibitem{Kapitza} P. Kapitza, Nature {\bf 141}, 74 (1938).
\bibitem{Allen} J.F. Allen and A.D. Misener,
Nature {\bf 141}, 75 (1938).
\bibitem{BEC} A. Einstein, 
%Quantentheorie des einatomigen idealen gas, 
Sitzungsberichte der Preussischen Akademie der Wissenschaften 261-267 (1924); A. Einstein,
 %Quantentheorie des einatomigen idealen gas 2, 
 Sitzungsberichte der Preussischen Akademie der Wissenschaften 3-14 (1925).
\bibitem{Keesom} W.H. Keesom and H.P. Keesom, Physica {\bf 2}, 557 (1935).
\bibitem{KP} R. Kronig and W. G. Penney, Proc. R. Soc. A {\bf 130}, 499 (1931).  
\bibitem{Tamm} I. Tamm, Z. Phys. {\bf 76}, 849 (1932).


\end{thebibliography}
\end{document}